\documentclass[twocolumn,trackchanges]{aastex62}
\usepackage{amsmath}
\usepackage{longtable}
\usepackage[para]{threeparttable}
\usepackage{natbib}
\usepackage{footmisc}

\newcommand{\hii}{\mbox{H~{\sc ii}~}}

\newcommand{\oiii}{\mbox{O~{\sc iii}~}}
\newcommand{\cii}{\mbox{C~{\sc ii}~}}
\newcommand{\siii}{\mbox{Si~{\sc ii}~}}

\newcommand{\siiv}{\mbox{Si~{\sc iv}~}}

\newcommand{\heiii}{\mbox{He~{\sc {ii+i}}~}}
\newcommand{\heii}{\mbox{He~{\sc ii}~}}
\newcommand{\hei}{\mbox{He~{\sc i}~}}
\newcommand{\nai}{\mbox{Na~{\sc i}~}}
\newcommand{\cai}{\mbox{Ca~{\sc i}~}}
\newcommand{\feii}{\mbox{Fe~{\sc ii}~}}
\newcommand{\niii}{\mbox{N~{\sc iii}~}}
\newcommand{\fei}{\mbox{Fe~{\sc i}~}}
\newcommand{\tii}{\mbox{Ti~{\sc i}~}}
\newcommand{\tiii}{\mbox{Ti~{\sc ii}~}}
\newcommand{\tio}{\mbox{Ti{O}~}}
\newcommand{\cri}{\mbox{Cr~{\sc i}~}}
\newcommand{\bai}{\mbox{Ba~{\sc i}~}}
\newcommand{\baii}{\mbox{Ba~{\sc ii}~}}
\newcommand{\mni}{\mbox{Mn~{\sc i}~}}

\received{...}
\revised{....}
\accepted{: August, 2018}

\shorttitle{The {\it Planck} cold clump G108.37-01.06}
\shortauthors{Dutta et al.}

\begin{document}
\nocite{*}
\title{THE {\it PLANCK} COLD CLUMP G108.37-01.06: A SITE OF COMPLEX INTERPLAY BETWEEN \hii REGIONS, YOUNG CLUSTERS AND FILAMENTS}

\correspondingauthor{Somnath Dutta}
\email{duttasomnath9@gmail.com}

\author[0000-0002-2338-4583]{Somnath Dutta}
\affil{Satyendra Nath Bose National Centre for Basic Sciences, \\
Block-JD, Sector-III, Salt Lake, Kolkata-700 106}

\author{Soumen Mondal}
\affiliation{Satyendra Nath Bose National Centre for Basic Sciences, \\
Block-JD, Sector-III, Salt Lake, Kolkata-700 106}

\author{Manash R Samal}
\affiliation{Physical Research Laboratory, Navrangpura, Ahmedabad, Gujarat 380009, India}

\author{Jessy Jose}
\affiliation{Indian Institute of Science Education and Research,\\
Rami Reddy Nagar, Karakambadi Road, Mangalam (P.O.), \\
Tirupati 517507, India}

\begin{abstract}
The {\it Planck} Galactic Cold Clumps (PGCCs) are the possible representations of the initial conditions and the very early stages of star formation. With an objective to understand better the star and star cluster formation, we probe the molecular cloud associated with PGCC G108.37-01.06 (hereafter, PG108.3),  which can be traced in a velocity range $-$57 to $-$51 km s$^{-1}$. The IPHAS images reveal H$\alpha$ emission at various locations around PG108.3, and optical spectroscopy of the bright sources in those zones of H$\alpha$ emission disclose two massive ionizing sources with spectral type O8-O9V and B1V. Using the radio continuum, we estimate ionizing gas parameters and find the dynamical ages of \hii regions associated with the massive stars in the range   0.5$-$0.75 Myr.  Based on the stellar surface density map constructed from the deep near-infrared CHFT observations, we find two prominent star clusters in PG108.3; of which, the cluster associated with \hii region S148 is moderately massive ($\sim$ 240 M$\sun$). A careful inspection of JCMT $^{13}$CO(3$-$2) molecular data exhibits that the massive cluster is associated with a number of filamentary structures. Several embedded young stellar objects (YSOs) are also identified in the PG108.3 along the length and junction of filaments. We find the evidence of velocity gradient along the length of the filaments. Along with kinematics of the filaments and the distribution of ionized, molecular gas and YSOs, we suggest that the cluster formation is most likely due to the longitudinal collapse of the most massive filament in PG108.3.
\end{abstract}


\section{Introduction} \label{sec:intro}
The formation of the star cluster is a topic of considerable interest since most stars in our Galaxy form in groups within clustered environments
\citep[e.g.,][]{2003ARA&A..41...57L}. Several environmental conditions can breed young clusters such as: i) fragmentation of the swept up matter in the shells of  the expanding  \hii regions \citep[e.g.][]{1977ApJ...214..725E,2008ApJ...688.1142K,2014A&A...566A.122S,2016ApJ...818...95L}, ii) external compression of pre-existing clumps by nearby massive stars \citep[e.g.][]{1994A&A...289..559L,2011MNRAS.415.1202C,2013MNRAS.432.3445J,2015ApJ...799...64D}, iii) matter sandwiched between bubbles \citep[e.g.][]{2001ApJ...553L.185Y,2011ApJ...738..156O}, and iv) at the collision point of  molecular clouds \citep[e.g.][]{2014ApJ...780...36F,2018PASJ..tmp....9W,2018PASJ..tmp...10H}, v) at the junction of converging filaments or hub of filamentary systems \citep[e.g.][]{2012A&A...540L..11S,2013ApJ...766..115K}. As molecular clouds are often comprised of \hii regions, bubbles, and dense filamentary structures \citep[][]{2009AJ....138..975B, 2013ApJ...779..113M, 2015A&A...581A...5S, 2016ApJS..226...13B, 2017ApJ...836...98J}, therefore, understanding what shapes the molecular clouds to dense and massive enough to form a young cluster is of great interest. Additionally, the initial cloud configuration decides the future location of cluster formation \citep[e.g.][]{2004ApJ...616..288B}. Therefore, the exact role of environment on the star and star cluster formation of a cloud can only be thoroughly understood by tracing various components of the interstellar medium (ISM) through multiwavelength observations.

{\it Planck} is the third generation mission to measure the anisotropy of the cosmic microwave background (CMB) at nine frequency bands from 30 to 857 GHz, with beam sizes ranging from 33$\arcmin$ to 5$\arcmin$ \citep[][]{2011A&A...536A..23P}. As the high-frequency channels of {\it Planck} cover the peak thermal emission of dust colder than 14 K, therefore, {\it Planck} images probe the coldest parts of the ISM.  In fact, using (sub)millimeter and millimeter wavelengths {\it Planck} have revealed an extremely cold population of dense molecular clumps, namely the {\it Planck} Galactic Cold Clumps \citep[PGCCs,][]{2016A&A...594A..28P}. The {\it Planck} Early Release Compact Source Catalogue \citep[ERCSC,][]{2011A&A...536A..13P} provides lists of positions and flux densities of compact sources at each of the nine {\it Planck} frequencies. The properties of PGCCs, however, are still not well known due to the lack of observations at high spatial resolution. Analysis of a part of the ERSSC suggests that depending upon distances, these cold sources trace a broad range of objects, from low-mass dense cores to giant molecular clouds \citep[][]{2018arXiv180503883Z}. High angular resolution {\it Herschel} and CO isotopologues follow-up observations have revealed that substantial fraction of such cold sources are filamentary \citep[][]{2012A&A...541A..12J,2017arXiv171109425J} and correspond to the very initial evolutionary stages of star formation \citep[][]{2012ApJ...756...76W,2013ApJS..209...37M,2016ApJS..224...43Z,2018ApJS..234...28L}. In particular, \citet[][]{2016A&A...591A.105Z} pointed out based on {\it Herschel} data, about 25\% of PGCCs near the Galactic mid-plane may be massive enough to form high-mass stars and star clusters.

In this work, we examine the molecular cloud and stellar content associated with PGCC G108.37-01.06 (hereafter PG108.3, for details see section \ref{sec:overview}) located at {\it l} = 108\degr.3710, {\it b}  = $-$1\degr.0649 ($\alpha_{2000}$ = 22$^{\rm h}$ 56$^{\rm m}$ 21$^{\rm s}$, $\delta_{2000}$ = +58\degr 30\arcmin 55\arcsec) using multiwavelength datasets. The large-scale CO morphology of the cloud is filamentary, it hosts a number of \hii regions \citep[][]{1985PASJ...37..345T}. Moreover, in the shallow near-infrared (NIR) 2MASS images, it appears to be a possible site of cluster formation. Therefore, PG108.3 is a good target to investigate the environmental effect on the formation of star and star clusters. Our aim is to use multiwavelength high-resolution observations to understand what dominates the star formation process in this planck cold clump. In this work, we made use of optical spectroscopic observations to identify the ionizing stars of the \hii regions. We studied ionized gas content and dynamical status of the \hii regions using radio continuum data. Using deep $H$, $K$ band data sets along with {\it Spitzer}-IRAC data, we identify and characterize the various young stellar objects (YSOs) and young clusters of the region. We examine the cold gas distributions and their kinematics using JCMT $^{13}$CO($3-2$) observations and probe their correlation with ionized gas, YSOs, and clusters. In the final step, we attempt to comprehend the star formation scenario of the complex. We 
organize our work as follows. In Section \ref{sec:overview} we present a brief overview of the PG108.3. Section \ref{sec:observations} describes our observational details. Section \ref{sec:results_of_ionized_gas} deals with various results on the ionized, stellar and young stellar, and gas component of the cloud. Section \ref{sec:understanding_of_star_and_cluster_formation} is devoted to our studies of star and cluster formation in the cloud based on the various results obtained in this analysis. We summarize our main findings in Section \ref{sec:conclusions}.

\begin{figure*}
 \centering
 \includegraphics[height=11.0cm, width=13.0cm]{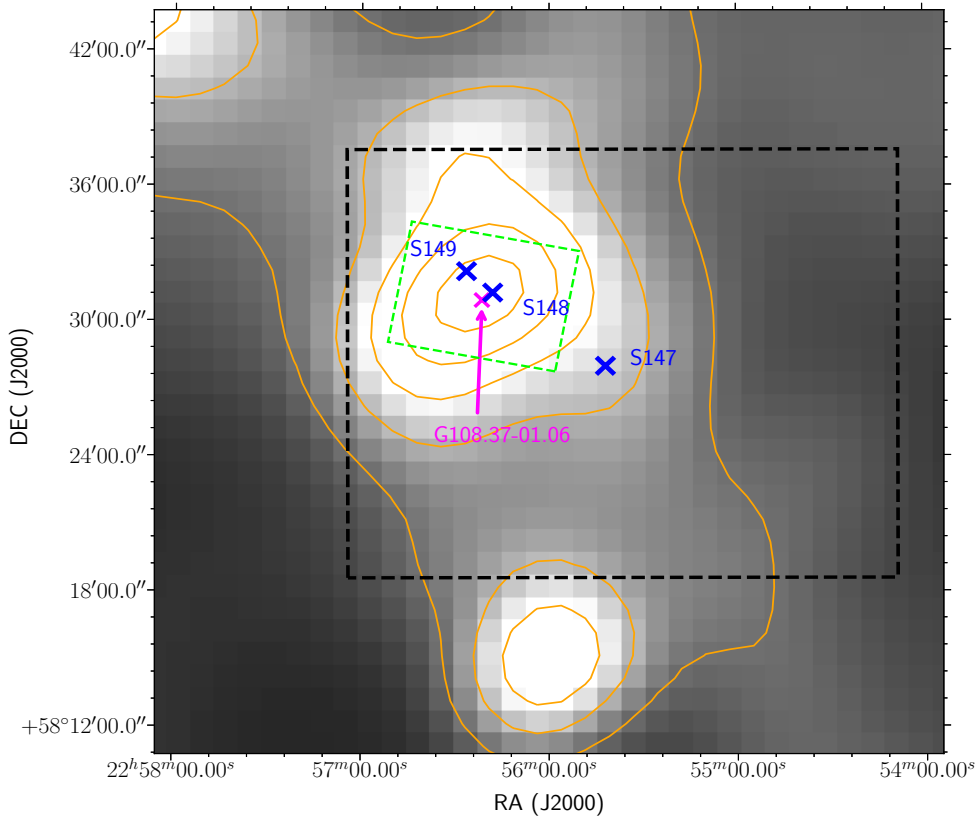}
  \caption{{\it Planck} 857 GHz view of the PG108.3. The simbad positions of three \hii regions are marked. The contour levels of background image are at 60.0, 80.0, 100.0, 130.0, 160.0 MJy sr$^{-1}$. Our FOV ($\sim$ 23$\arcmin$ $\times$ 19$\arcmin$) in CFHT WIRCam is shown in black box, whereas green box is the region covered by {\it Spitzer}-IRAC. See text for details.} 
  \label{fig:planck}  
\end{figure*}

  \section{Overview of the PG108.3 complex} \label{sec:overview}
 Figure \ref{fig:planck} shows the location of the bright clump PG108.3, which is located in Perseus arm of our Milky Way. Three \hii regions \citep[S149, S148, S147;][]{1959ApJS....4..257S} are projected in the field of PG108.3 complex. The position of S148 nearly coincides with PG108.3. In the direction of PG108.3, the radial velocity peaks of the molecular gas were found to be $-$54.8, $-$55.8, $-$49.4 km s$^{-1}$ for S149, S148 and S147, respectively \citep[e.g.,][]{1975A&A....40..347B,1975A&A....39..481G,1978A&A....66....1C,1979ApJ...231L.115B,1985PASJ...37..345T,2011AJ....141..123A}. Recent observations on molecular gas around PG108.3 were performed by \citet[][]{2011AJ....141..123A}. Using James Clerk Maxwell Telescope (JCMT) observations in $^{13}$CO(2-1) and  $^{12}$CO(2$-$1) with a velocity resolution 0.04 km s$^{-1}$ for both, several cold clumps were detected in the mass range 190 to 1400 $M\sun$ \citep[][]{2011AJ....141..123A} assuming the distance of the region S148-S149 as 5.6 $\pm$ 0.6 kpc. Such massive clumps are the possible location of on-going or future formation of stellar cluster. However, we note that the mass of CO clumps as estimated by  \citet[][]{2011AJ....141..123A} is likely an overestimation  as they have adopted a distance 5.6 kpc for the region, while from the recent works and as well as from this work we suggest that 3.3 kpc is likely  the more appropriate distance to the region (see also section \ref{sec:distance_mass}). \citet[][]{1985PASJ...37..345T} suggests that the group S149-147 is possibly attached with the group of \hii regions S153-152 from their observations in the $^{12}$CO(1$-$0) and $^{13}$CO(1$-$0) molecular lines. The group S153-152 is located towards the north of S149-147, with an average radial velocity of $-50$ km s$^{-1}$.

\section{Observations and Data Reduction} \label{sec:observations}

\subsection{Spectroscopic Observations}
     With an aim of investigating the probable massive ionizing sources, we carried out optical spectroscopic observations of 5 bright sources in the optically nebulous region around PG108.3 (see section~\ref{sec:results_of_ionized_gas}) using HFOSC instrument on 2m Himalayan Chandra Telescope (HCT), India. The observations were performed with Grism 7 (3800- 6840 {\AA}) with a resolving power of 1200 and Grism 8 (5800- 8350 {\AA}) with a resolving power of 2190. Proper lamp and bias frames were taken immediately after the target observations. The spectroscopic standard star G191B2B \citep[][]{1990AJ.....99.1621O} was also observed with an exposure time of 600s for the flux calibration. 

    After bias subtraction and flat field correction, the one-dimensional spectra were extracted using the optimal extraction method with APALL task in IRAF\footnote{Image Reduction and Analysis Facility (IRAF) is distributed by National Optical Astronomy Observatories (NOAO), USA (http://iraf.noao.edu/)} and wavelength calibrated using the FeAr arc lamp observations. The spectra were also corrected for the instrumental response using the sensitivity function generated from the standard star observations.

\subsection{NIR Imaging}\label{sec:nir_imaging}
Deep NIR observations around PG108.3 (centered on $\alpha_{2000}$ = $22^h55^m36^s$ $\delta_{2000}$ = $+58^028^m06^s$) were obtained in $H$ (1.63 $\mu$m), $K$ (2.14 $\mu$m) bands covering a field-of-view (FOV) $\sim$ 23$\arcmin$ $\times$ 19$\arcmin$ from CFHT data archive\footnote{http://www.cadc-ccda.hia-iha.nrc-cnrc.gc.ca/en/cfht/} observed using WIRCam camera of the CFHT 3.6m telescope during 2007 July 31. In this observing set up, each pixel corresponds to 0.3$\arcsec$ and yields a FOV $\sim$ 20$\arcmin$ $\times$ 20$\arcmin$ in each exposure. The average FWHM during observations was $\sim$ 0.69$\arcsec$. 
The $H$ and $K$  images were observed in several dithered position with four frames in each position. Each of the $H$ and $K$ frames were taken with the  exposure time of 14.80 s and 23.8 s resulting total effective exposures 300 s and 2000 s, respectively.

The data were reduced using Interactive Data Language (IDL) based reduction pipeline-SIMPLE Imaging and Mosaicking PipeLine \citep[SIMPLE;][]{2010ApJS..187..251W}. This pipeline generates sky-flat from median combining of the dithered images. It provides good treatment to sky background fluctuation and minimized artifacts from bright objects. Absolute astrometry solution was obtained with 2MASS reference catalog. The images were calibrated with the 2MASS catalog. In order to avoid nonlinearity due to saturated stars in WIRCam bright end, we considered only stars in 13-14.5 mag for $K$ and 13-15 mag for $H$ with good photometric accuracy (err $\le$ 0.1) for the calibration of photometric images. The identification of point sources was performed with the DAOFIND task in IRAF. Following \citet[][]{1987PASP...99..191S}, roundness limits of $-1$ to $+1$ and sharpness limits of $0.2$ to $+1$ were applied to eliminate bad pixels brightness enhancements and the extended sources such as background galaxies from the point source catalog. The photometry on the images was performed with PSF algorithm of DAOPHOT package \citep[][]{1992ASPC...25..297S}.

  To avoid the inclusion of WIRCam saturated sources we replaced all the sources in our catalog having 2MASS magnitudes $H$ $\leq$ 13.75 mag and $K$ $\leq$ 13.5 mag. We do not have WIRCam $J$ band observations. We include all available 2MASS $J$ magnitude of the detected sources. Thus, our final catalog is sensitive down to $\sim$ 20.0 mag and $\sim$ 20.5 mag in $H$ and $K$-band, respectively, with good photometric accuracy (error $\le$ 0.1 mag).

\subsection{{\it Spitzer}-IRAC data} 

The {\it Spitzer}-IRAC observations in 3.6 and 4.5 $\mu$m bands were available in the {\it Spitzer} archive program (Program ID: 30734 ; PI: Figer, Donald F) for the central part (marked in the green box in Figure~\ref{fig:planck}) towards the PG108.3. The basic calibrated Data (version S18.18.0) were downloaded from Spitzer archive\footnote{http://archive.spitzer.caltech.edu/}. The raw data were processed and the final mosaic frames were created using MOPEX (version 18.5.0). We performed point response function (PRF) fitting method using APEX tool provided by {\it Spitzer} Science center on the mosaic {\it Spitzer}-IRAC images to extract the magnitudes of point sources. The objects with good photometric accuracy (error $\le$ 0.2 mag) were utilized for this analysis.

\begin{figure}
 \centering
 \includegraphics[scale=0.37]{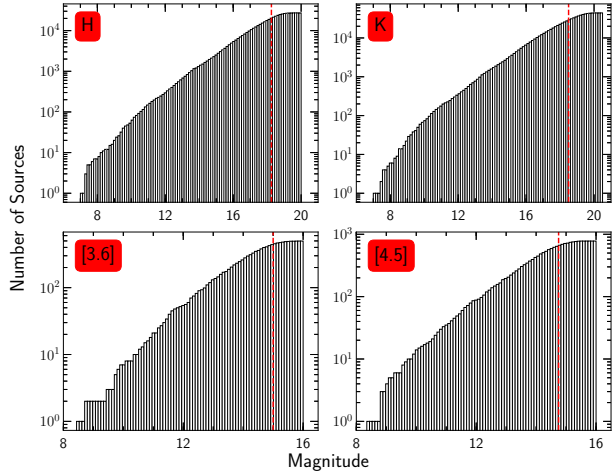}
  \caption{The cumulative distribution of the sources detected in various wavebands. The vertical red lines indicate the 90\% completeness limit of the corresponding wave bands. See text for details.}  
  \label{fig:completeness}  
\end{figure}

   The completeness limits at various bands were estimated from histogram turn over method \citep[e.g.,][]{2013A&A...552A..14O,2015A&A...581A...5S}. Figure \ref{fig:completeness} displays the cumulative logarithmic distribution of sources at different wavebands. We considered $\sim$ 90\% completeness of our data at the magnitudes at which the histograms deviate from linear distribution and are indicated by vertical lines in Figure \ref{fig:completeness}. With the above approach, our photometry is $\sim$ 90\% complete down to $H$ = 18.25 mag, $K$ = 18.50 mag, [3.6] = 15.00 mag and [4.5] = 14.75 mag, respectively. However, in the high nebulous region, we expect 2-5\% brighter completeness limits with respect to present estimated values.

\subsection{\it Other ancillary datasets}      

Planck: The {\it Planck  satellite} surveyed the  entire sky in nine frequency wavebands during 2009-2013 \citep[][]{2011A&A...536A..23P} with the High Frequency Instrument (HFI; 857,  545,  353,  217,  143  and  100 GHz) impacting the effective angular resolution ranging from 5$\arcmin$ to 9.6$\arcmin$ \citep[][]{2010A&A...520A...9L,2011A&A...536A...4P}.

JCMT: The James Clerk Maxwell Telescope (JCMT) is operated by the  Joint Astronomy Centre on behalf of the Science and Technology Facilities Council of the United Kingdom, the Netherlands Organisation for Scientific Research, and the National Research Council of Canada. Heterodyne Array Receiver Program (HARP) installed on the JCMT operates in the submillimetre frequency bands (325  to  375  GHz) allowing simultaneous observations of multiple lines within an intermediate frequency of 5 GHz \citep{2009MNRAS.399.1026B}. This study utilized  $^{13}$CO(3$-$2) (330.58796 GHz) observations (Program ID: M10BU08)  available at JCMT data archive\footnote{http://www.cadc-ccda.hia-iha.nrc-cnrc.gc.ca/en/jcmt/}. This dataset has a velocity resolution of 0.05 km s$^{-1}$ and rms level is 0.16 K. The JCMT beam size at 330.58796 GHz is 15.2 arcsec; and the main beam efficiency ($\eta$mb) is 0.61.

      We also obtained the Submillimetre Common-User Bolometer Array 2 (SCUBA-2) 850 $\mu$m observations from the JCMT archive to trace the dust continuum emission in the star-forming region \citep[SFR;][]{2013MNRAS.430.2534D}. The beam FWHM of SCUBA-2 at 850 $\mu$m is 13 arcsec with a mean rms noise level of $\sim$ 2.9 mJy beam$^{-1}$.

IRAS: The Infrared Astronomical Satellite (IRAS) was the first-ever space telescope to perform a survey of the entire night sky at 12, 25, 60, and 100 $\mu$m wavelengths.  Using the High Resolution Image Restoration \citep[HIRES;][]{2004AJ....127.3235S} image reconstruction technique,  a spatial resolution ranging from 30$\arcsec$ to 1.5$\arcmin$ can be achieved for the IRAS images. For this analysis, we obtained the IRAS images.

IPHAS: INT Photometric H$\alpha$ Survey (IPHAS\footnote{http://www.iphas.org/data.shtml}) is a photometry survey in the Northern Milky Way in visible light (H$\alpha$, r, i) using wide-field camera on the 2.5-m Isaac Newton Telescope (INT) in La Palma \citep[][]{2005MNRAS.362..753D,2008MNRAS.388...89G,2014MNRAS.444.3230B}. We used  H$\alpha$ images from the IPHAS survey to trace the ionized gas.

 Gaia DR2 data: We also utilized the Gaia Data Release 2 \citep[][]{2018arXiv180409365G} from the European Space Agency (ESA) mission Gaia \citep[][]{2016A&A...595A...1G} to estimate the distance of the cloud complex.

WISE: The WISE survey  provides photometry at four wavelengths 3.4, 4.6, 12 and 22 $\mu$m with an angular resolution of 6.1\arcsec, 6.4\arcsec, 6.5\arcsec, 12.0\arcsec, respectively \citep[][]{2012yCat.2311....0C}.  We used WISE images for our analyses.

Radio continuum emission: We made use of National Radio Astronomy Observatory (NRAO) VLA Sky Survey (NVSS) map at 1.4 GHz with beam size 45$\arcsec$  \citep[NVSS;][]{1998AJ....115.1693C} to trace properties of ionized gas around PG108.3.

\section{Ionzied, stellar and molecular components of the clump} \label{sec:results_of_ionized_gas}

\subsection{Optically Visible \hii regions in PG108.3} \label{sec:distance_mass}
\begin{figure*}[t!]
     \centering
 \includegraphics[width=17.0cm, height=6.0cm]{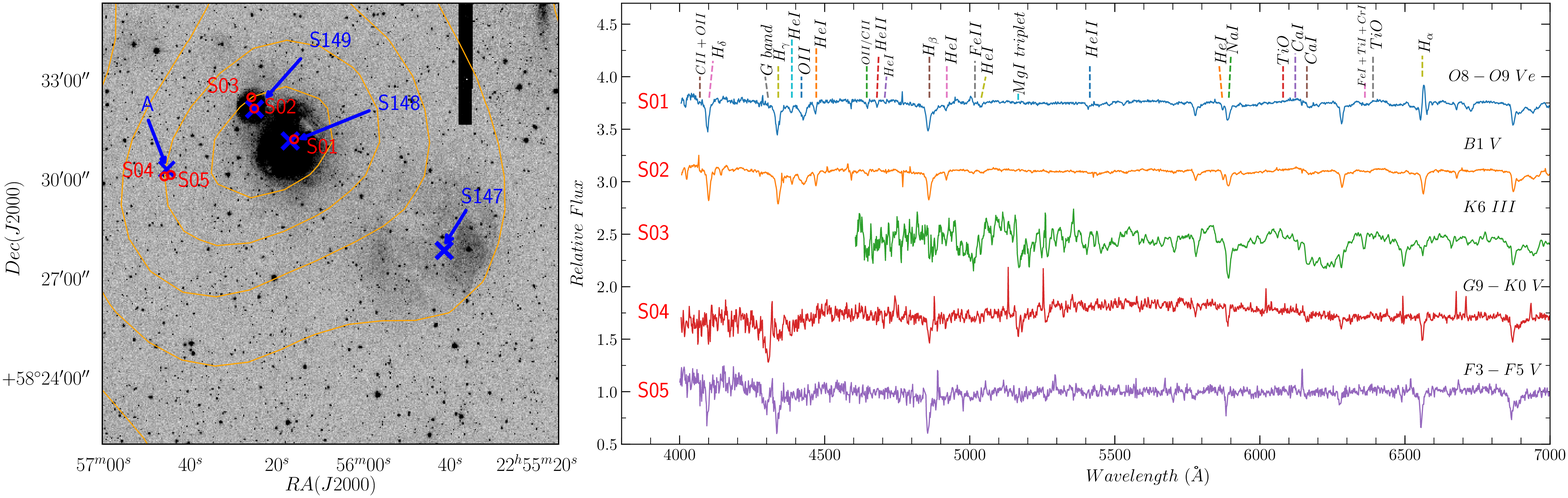}
  \caption{({\it left panel}) Optical H$\alpha$ emission is displayed. The spectroscopically observed stars are marked. The projected \hii regions (S149, S148 \& S147) including subregion `A' (see section \ref{sec:surface_density}) are also shown. The contours of {\it Planck} 857 GHz continuum emission have the same meaning as in Figure \ref{fig:planck}. ({\it right panel}) The wavelength and flux calibrated spectra of the observed sources. The perceptible lines are marked on the top (see text for details).}  
  \label{fig:halpha_spectra} 
     \label{fig:opt_spec} 
\end{figure*}

 In the left panel of Figure \ref{fig:halpha_spectra}, the optical H$\alpha$ emission is visible towards S149, S148 and S147. In addition, we find patchy faint H$\alpha$  nebulosity in the south-eastern direction of S148 whose location corresponds to a faint NIR cluster visible in the $K$-band image (see Section \ref{sec:surface_density}), thus possibly a \hii region is in the process of emerging. We search for ionizing sources of these \hii regions, particularly sources lying around the peak of the clump as seen at 857 GHz. In the Figure \ref{fig:halpha_spectra} (left panel), the bright sources selected for spectroscopic observations are marked. The coordinates and optical magnitudes of these sources are given in Table~\ref{tab:spec}, and the  flux calibrated, normalized spectra are shown in Figure~\ref{fig:halpha_spectra} ({\it right panel}).

\subsubsection{Exciting stars and distance to the \hii regions} \label{sec:spectra_massive_stars}
The observed spectra were classified in comparison with the different spectral libraries available in the literature \citep[][]{1990PASP..102..379W,1984ApJS...56..257J,1993PASP..105..693T,1995AJ....109.1379A}. First, we determined a specific spectral range from strong conspicuous features \citep[see][for example]{2015MNRAS.454.3597D}. The line strengths of program spectra were echoed with the literature spectra within the above specified spectral range. Finally, each source was visually compared to the standard library spectra. Photometric and spectroscopic details of all the  5 sources are given in Table~\ref{tab:spec}. Based on low-resolution spectra of our targets, an uncertainty of $\pm$1 in the sub$-$class estimation is expected. 
On the basis of the low$-$resolution spectroscopy of early type stars, it is difficult to distinguish the luminosity class  between supergiants, giants, dwarfs and pre$-$main sequence (PMS) stars. The spectral features are described as follows:

S01: The star S01 appears to lie within strong nebulosity of S148. It is classifiable from hydrogen and \heii, \hei lines. Presence of \heii lines ($\lambda\lambda$ 4686, 5411) and \heiii ($\lambda\lambda$ 4026 \AA) limits the spectral type to O-type. Prominent absorption bands occur at hydrogen ($\lambda\lambda$ 3970, 4101,  4340,  4861,  6563 \AA), \hei ($\lambda\lambda$ 4144, 4387, 4471, 4713 \AA), \siiv ($\lambda\lambda$ 4089, 4116 \AA), blended \cii + \oiii ($\lambda\lambda$ 4070, 4650 \AA), \niii ($\lambda\lambda$ 4379, 4641 \AA). Moderate nitrogen enhancement indicate a later O-type spectrum. The line ratios of \heii ($\lambda\lambda$ 4686)/\hei  ($\lambda\lambda$ 4713), \siiv ($\lambda\lambda$ 4089)/\hei ($\lambda\lambda$ 4144, 4387, 4471, 4713 \AA) and \siiv ($\lambda\lambda$ 4116)/\hei ($\lambda\lambda$ 4144) are revealing the star more than towards O8-O9 V. A P-cygni profile is seen at $H\alpha$-band, indicating strong accretion activity of the star.

S02: This star is located within high nebulous region of S149. The absence of prominent \hii lines and presence of \hei ($\lambda\lambda$ 4009, 4026, 4121, 4144, 4387, 4471, 4713, 5876 \AA) targets the star S02 within a spectral range in B0.5 to A0-type. The spectrum shows absorption at \niii ($\lambda\lambda$ 4511-15, 4641 \AA), \cii ($\lambda\lambda$ 4216 \AA), \oiii ($\lambda\lambda$ 4415-4417 \AA) \cii + \oiii ($\lambda\lambda$ 4070, 4650 \AA), \siiv ($\lambda\lambda$ 4089, 4116, 4128  \AA), \siii ($\lambda\lambda$ 4552-4568 \AA). The ratio of \siii ($\lambda\lambda$ 4552) / \siiv ($\lambda\lambda$ 4128 \AA), and comparison of \hei strength suggests the S02 is more similar to B1 V spectrum.

S03:  The main conspicuous features obtained in the spectrum of S03 are absorption lines \fei at $\lambda\lambda$ 7749, 7834 \AA, \cai at $\lambda\lambda$ 6162 \AA, $H\alpha$, blended \fei, \tii, \cri at $\lambda\lambda$ 6362 \AA, blended \bai, \fei, \cai, \mni, \tii and \tiii at $\lambda\lambda$ 6497 \AA, \tio band at $\lambda\lambda$ 5847-6058, 6080-6390  \AA. These stellar features identified in this spectrum of S03 explain its trends, both with spectral type and luminosity class as K6 III.

S04:  The presence of G band ($\lambda\lambda$ 4300) in its spectrum indicating it as after F2 type. The discernible absorption features are \cai ($\lambda\lambda$ 4226), \fei ($\lambda\lambda$ 4326), \cai ($\lambda\lambda$ 4226), \nai ($\lambda\lambda$ 5890), ( \baii, \fei, \cai ($\lambda\lambda $ 6497), $H\alpha$, \cai ($\lambda\lambda$ 8498, 8542, 8662), CaH ($\lambda\lambda$ 6386). Literature comparison of these features along with line strength of $H\alpha$ and \nai makes S04 more similar to G9-K0 V.

S05: Based on the late-type absorption features at G band ($\lambda\lambda$ 4300), \fei ($\lambda\lambda$ 4383), \mni ($\lambda\lambda$ 4030), \fei ($\lambda\lambda$ 4046), \cai ($\lambda\lambda$ 6162), \feii ($\lambda\lambda$ 6242, 6456), \fei+ \cai($\lambda\lambda$ 6497) in this object, we agree with the spectral range F3-F5 V.

\begin{table*}

\caption{Details of the spectroscopically observed stars}
\small
\centering
\label{tab:spec}
\begin{tabular}{ccccccccccc}

\hline 
\multicolumn{1}{c}{ID} & \multicolumn{1}{c}{RA (J2000)} & \multicolumn{1}{c}{Dec (J2000)} & \multicolumn{1}{c}{Spectral}  & \multicolumn{1}{c}{$J$} &  \multicolumn{1}{c}{$J-H$} & \multicolumn{1}{c}{Spectroscopic} & \multicolumn{1}{c}{Spectroscopic} & \multicolumn{1}{c}{Gaia} & \multicolumn{1}{c}{Remarks} \\ 
\multicolumn{1}{c}{} & \multicolumn{1}{c}{(h:m:s)} & \multicolumn{1}{c}{(d:m:s)} & \multicolumn{1}{c}{Type} & \multicolumn{1}{c}{(mag)} &  \multicolumn{1}{c}{(mag)} & \multicolumn{1}{c}{$A_V$ (mag)} & \multicolumn{1}{c}{distance (pc)} & \multicolumn{1}{c}{distance (pc)}& \multicolumn{1}{c}{} \\ \hline
\hline

S01 & 22:56:17.17  &  +58:31:17.98  & O8-O9 Ve & 10.572 & 0.348 & 4.65 $\pm$ 0.22  & 3600 $\pm$ 330 & $...$            & S148\\
S02 & 22:56:26.52  &  +58:32:12.76  & B1 V     & 11.179 & 0.205 & 3.21 $\pm$ 0.16  & 3590 $\pm$ 250 & 3400 $\pm$ 324   & S149\\
S03 & 22:56:27.24  &  +58:32:33.13  & K6 III   & 6.100  & 1.432 & $...$            & $...$          & $...$            & background\\
S04 & 22:56:47.07  &  +58:30:07.07  & G9-K0 V  & 13.832 & 0.428 & 0.573 $\pm$ 0.35 & 810 $\pm$ 180  & 1328 $\pm$ 80    & foreground\\
S05 & 22:56:45.600 &  +58:30:09.90  & F3-F5 V  & 14.410 & 0.437 & 0.812 $\pm$ 0.25 & 1920 $\pm$ 270  & 2137 $\pm$ 217   & foreground\\

\hline\end{tabular}
\begin{tablenotes}\footnotesize
\end{tablenotes}
\end{table*}

\begin{figure}
\includegraphics[width=8.0 cm,angle=0]{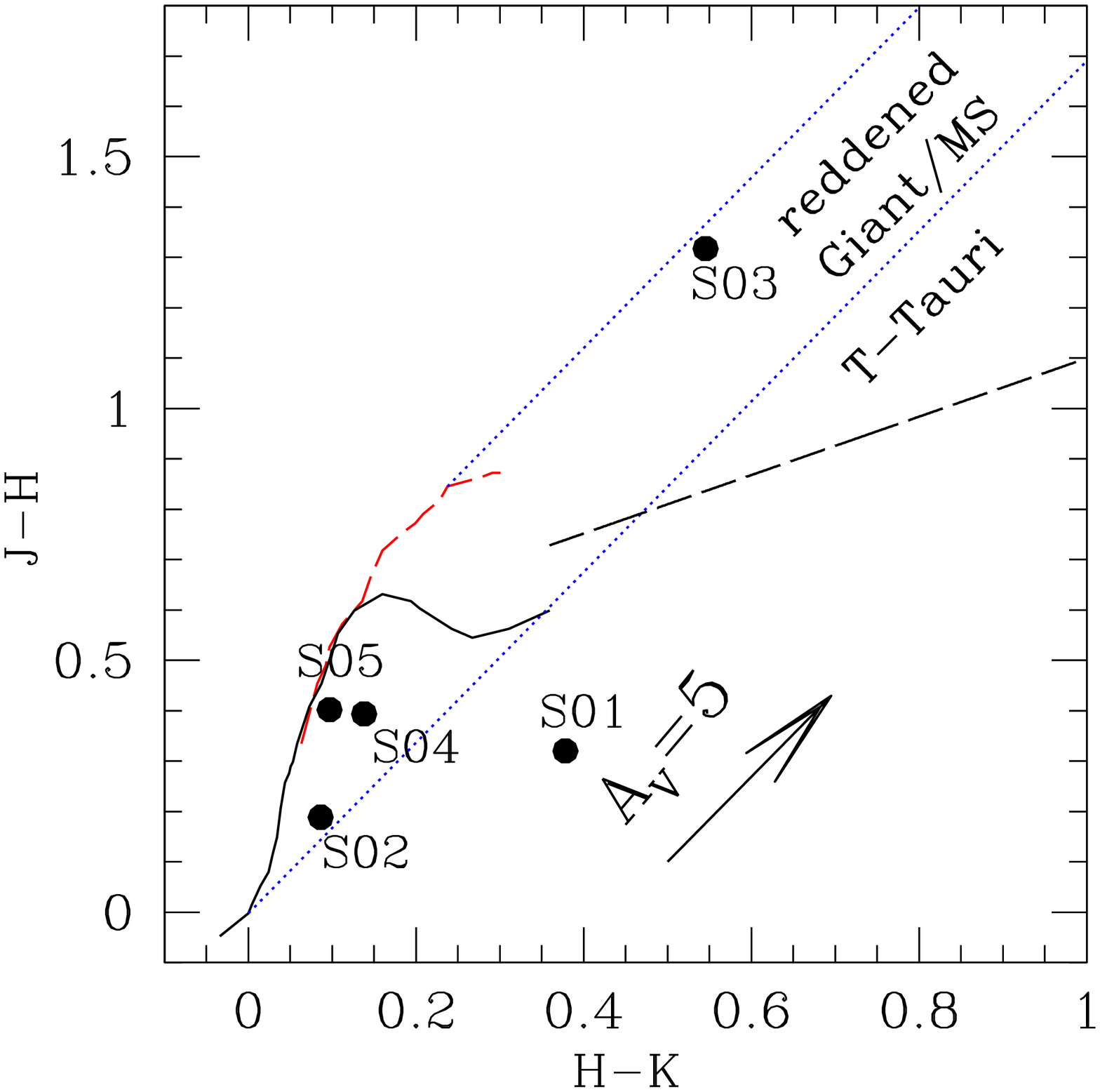}
  \caption{($J-H$)/($H-K$) CC diagram of the spectroscopically observed stars in PG108.3. Their ids are as given in Table \ref{tab:spec}. The sequences for dwarfs (solid black) and giants (red dashed line) are taken from \citet[][]{1988PASP..100.1134B}. The long dashed black line represents the CTTSs locus \citep[][]{1997AJ....114..288M}. The dotted blue lines represent the reddening vectors \citep[][]{1981ApJ...249..481C}. A representative reddening vector of A$_V$ = 5 mag is also shown.}  
  \label{fig:spec_jhk}  
\end{figure}

All the spectroscopically observed stars are shown in ($J-H$)/($H-K$) colour-colour (CC) diagram in Figure~\ref{fig:spec_jhk}. The 2MASS and WIRCam magnitudes and colours were converted to the CIT system using the relations given by \citet[][]{2001AJ....121.2851C}. The sources towards the right of the right reddening vector at `T-Tauri' location are assumed to be the location of classical T-Tauri stars (CTTSs) \citep[e.g.,][and references therein]{2018MNRAS.476.2813D}. 
The star S01 is located right to the locus of massive main sequence (MS) star \citep[i.e in the zone of Herbig Be stars,][]{1992ApJ...397..613H}, possibly due to excess emission from the circumstellar disk. The star S02 is close to the locus of massive MS stars, which is a moderately extinct massive star. The ids S04 and S05  close to MS branch, which supports our spectral classification. The star S03 appears to be a reddened giant star, nonetheless, its high uncertainty ($\sim$ 0.4 mag) in the 2MASS $J$-mag makes it difficult to draw any conclusive remark.

Following the prescription of \citet[][]{2015MNRAS.454.3597D}, we estimated the visual extinction ($A_V$) to each observed source from their observed photometry and intrinsic colours adopted from \citet[][]{2013ApJS..208....9P} according to their spectral classification. The distances were measured from their absolute magnitudes and estimated  $A_V$, which are presented in Table~\ref{tab:spec}. The membership of the candidates was investigated based on their spectral types and distance. The distance for star ids S01 and S02 are  estimated to be 3600 $\pm$ 330 and 3590 $\pm$ 250 pc, respectively. Our estimated spectral classifications and distances for S01 and S02 agree well with the published values of S152-S153 \hii regions in the literature \citep[][]{2011A&A...535A...8R}, which are likely part of the same cloud in which the S149-S148 \hii regions are associated \citep[see Section \ref{sec:overview} and also][for details]{1985PASJ...37..345T}. Considering their distances, S04 and S05 seem to be foreground stars. The distance of star S03 was not estimated since we were unable to specify the luminosity class of the star from low-resolution spectra. We considered S03 as background giant stars after visual inspection of multiwavelength images and its location on ($J-H$)/($H-K$) CC diagram.

 We also searched for Gaia DR2 parallax for bright sources ($G$ $<$ 16 mag) in the nebulous region, with good relative parallax uncertainties ($\varpi$/$\sigma_\varpi$ $>$ 5) to compute distances ($d$ = $1/\varpi$) within 20 \% uncertainty limit \citep[][]{2018arXiv180409376L}. Under the above criterion, we find the distance of S02 (d $\sim$ 3.4 $\pm$ 0.32 kpc, see Table \ref{tab:spec}) remain within error bars of spectroscopic distance. We also notice, the distances of S04 \& S05 are close to the spectroscopic measurements. We reveal one source  ($\alpha_{2000}$ = $22^{\rm h}$ 56$^{\rm m}$ 17.46$^{\rm s}$, $\delta_{2000}$ = +58\degr 31\arcmin 17.5\arcsec) very close to S01 (within 0.04 pc), thus likely member of S148 cluster. The Gaia distance of the source is 3.1 $\pm$ 0.22 kpc. At around the faint nebulosity ($\alpha_{2000}$ = $22^{\rm h}$ 56$^{\rm m}$ 46$^{\rm s}$, $\delta_{2000}$ = +58\degr 30\arcmin 26\arcsec) in the western part, three sources are found to persist above criteria of Gaia DR2, of which two (S04 \& S05) are classified as foreground from spectrophotometry, another source ($\alpha_{2000}$ = $22^{\rm h}$ 56$^{\rm m}$ 45.35$^{\rm s}$, $\delta_{2000}$ = +58\degr 30\arcmin 24\arcsec) is estimated to be located at $\sim$ 3.55 $\pm$ 0.52 kpc, which is also classified as a class II YSO in  PG108.3 (discussed in section \ref{young_star_search}). Thus, based on Gaia DR2 datasets, we adopt an average distance of 3.3  $\pm$ 0.22 kpc of PG108.3 and associated \hii regions, as the Gaia measurements are likely to be more robust compared to our spectrophotmetric observations.

\subsubsection{Properties of ionized gas} \label{sec:Properties_of_ionized_gas}
In this section, we investigate the ionized gas associated with the \hii regions using the NVSS 1.4 GHz radio continuum map  (the NVSS  contours are shown in Figure \ref{fig:iras}). The number of Lyman continuum photons ($N_{Ly}$) can be computed from the observed integrated flux density using the following the equation \citep[see][]{2005A&A...433..205M}: 
\begin{equation}\label{equation:Lymann}
N_{Ly} = \frac{7.603\times 10^{46}\mathrm{ s}^{-1}}{b(\nu,T_e)}\left(\frac{S_\nu}{\mathrm{Jy}}\right)\left(\frac{T_e}{10^4\mathrm{
    K}}\right)^{-0.33}\left(\frac{D}{\mathrm{kpc}}\right)^{2}
\end{equation}
where  $S_\nu$ is the radio continuum integrated flux density at frequency $\nu$, $T_e$ is the electron temperature, $\theta_D$ is the angular diameter of the source, $D$ the distance from the Sun. The values of $S_\nu$ were adopted from NVSS 1.4 GHz catalog\footnote{https://www.cv.nrao.edu/nvss/}. The electron temperature $T_e$ were presumed from the galactocentric distance of the source \citep[R$_{gal}$ = 11.71 and 11.72 kpc for S148 and S149, respectively;][]{2000MNRAS.311..317C} following the relation given by \citet[][]{2014A&A...568A...4T}: 
\begin{equation}
T_e = 278\mathrm{K}\left(\frac{\mathrm{R_{gal}}}{\mathrm{kpc}}\right) + 6080\mathrm{K}
\end{equation}

Following \citet[][]{1978A&A....70..411P}, the rms electron density ($n_e$) can be obtained from the radio continuum map at 1.4 GHz considering spherical geometry of \hii regions: 
\begin{equation}\label{eq_flux}
\begin{split}
n_{e}  = \frac{4.092\times 10^{5}\mathrm{ cm}^{-3}}{\sqrt{b(\nu,T_e)}}
\left(\frac{S_\nu}{\mathrm{Jy}}\right)^{0.5}\left(\frac{T_e}{10^4\mathrm{
    K}}\right)^{0.25} \\ \left(\frac{D}{\mathrm{kpc}}\right)^{-0.5} \left(\frac{\theta_D}{arcsec}\right)^{-1.5}
\end{split}
\end{equation}
Where, $S_\nu$, $T_e$, D are same as defined for Equation \ref{equation:Lymann}, $\theta_D$ is the angular diameter in arcsec and  
\begin{equation}
b(\nu,T_e) = 1+0.3195\log\left(\frac{T_e}{10^4\mathrm{K}}\right)-0.2130\log\left(\frac{\nu}{\mathrm{GHz}}\right)
\end{equation}

The pressure from the ionized gas  is then estimated following \citet[][]{2014A&A...568A...4T}:
\begin{equation}
\mathrm{P_{e}} = \mathrm{2}n_{e}\mathrm{k_bT_e}
\end{equation}

 Where $k_b$ is the Boltzmann's constant. The measured values of S$_\nu$, size, $N_{Ly}$ (or $\log$($N_{Ly}$), $n_{e}$, $P_{e}$ are tabulated in Table \ref{tab:ionizing_flux}. Using the above formalism, we found that the values of $\log$($N_{Ly}$) were equivalent to the Lyman continuum photons  coming from a star of spectral type  O9-O9.5 V and B0.5-B1 V for S01  and S02, respectively \citep[see][]{1973AJ.....78..929P,2005A&A...433..205M}. These estimation are consistent with the spectral type estimated using optical spectra (see section \ref{sec:spectra_massive_stars}). It is always worthwhile to compare the present \hii region sample to nearby other \hii regions \citep[e.g., from][]{1989ApJS...69..831W,1994ApJS...91..659K,2003A&A...407..957M}. A comparison of size-density parameters to such large sample displays that S148 is now evolved to classical \hii region \citep[see also Figure 4 of][]{2005A&A...433..205M}, whereas S149 is in the transition from compact to classical \hii region.

\subsubsection{Evolutionary status of the \hii regions} 
\begin{figure}
\centering
\includegraphics[width=9cm, height=8cm, angle=0]{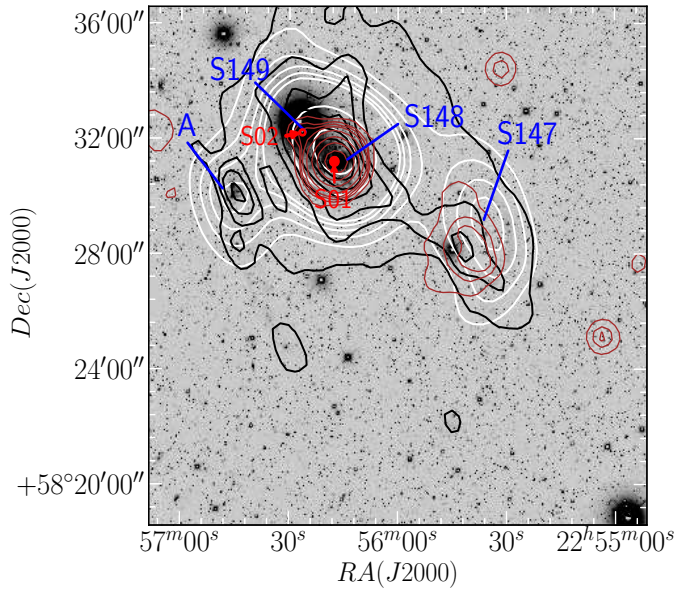}
  \caption{The locations of two massive sources (S01, S02) are shown in WIRCam $K$-band image. The warm dust contours (black) at 25 $\mu$m and cold dust contours (white) at 100 $\mu$m  are taken from IRAS images. The 25 $\mu$m contour levels are at 2\%, 3\%, 7\%, 15\%, 30\%, 50\%, 75\% of the peak value 670 MJy sr$^{-1}$ and the 100 $\mu$m contours are at 7\%, 9\%, 11\%, 13\%, 15\%, 35\%, 55\%, 75\% of the peak value 2650 MJy sr$^{-1}$. The 1.4 GHz contour (brown) are over plotted, contour levels are at 1.5, 4.0, 6.0, 10.0, 20.0, 50.0, 100.0, 200.0 mJy beam$^{-1}$. The locations of subregions (\hii regions: S149, S148, S147) including cluster A (see section \ref{sec:surface_density}) are marked.}  
  \label{fig:iras}  
\end{figure}

In this section, we examine the evolutionary stages of the \hii regions based on the 3D turbulent simulation of \citep[][]{2014A&A...568A...4T}, who 
followed the evolution of \hii regions in turbulent molecular clouds under the assumption that molecular clouds follow 
Larson's size-width relationship. We estimate the dynamical ages of the \hii regions by comparing the observed ionizing gas parameters (see Table \ref{tab:ionizing_flux}) with the pressure-size tracks of \citet[][]{2014A&A...568A...4T} and find that the dynamical ages of S148 and S149 are likely in the range of 0.5$-$0.75 Myr. However, we note that these ages are likely lower limit
as the method of \citet[][]{2014A&A...568A...4T} is more appropriate for large classical \hii regions where effects of magnetic field and gravity 
are less important. Given the fact that the \hii regions are still surrounded by a significant amount of gas and dust (see Figure \ref{fig:iras}, \ref{fig:spatial:ysos:integrated}), we consider 1 Myr 
as the likely age of the \hii regions. Here, we did not measure the age of S147, as the radio emission does not seem to be representing the entire \hii region (possibly due to lower sensitivity of the NVSS survey and/or missing flux problems of the  radio emission due to interferometric nature of radio observations) when compared to the emission seen in  H$\alpha$ or the outer extent of the \hii region cavity seen in WISE 12 $\mu$m emission.  
Figure \ref{fig:iras} shows the distribution of warm and relatively colder dust in  PG108.3 at 25 $\mu$m and 100 $\mu m$, respectively. The NVSS 1.4 GHz contours are also overplotted to trace the ionized gas in PG108.3. The peaks of cold and warm dust, along with the peak of molecular gas at CO 
\citep[see Figure 2b of][]{1985PASJ...37..345T} suggest that the gas-dust coupling is more efficient at the location of S148-S149 in contrast to S147 which is normally the case for regions of recent star formation \citep[e.g.,][]{2004A&A...415.1039O,2007ApJ...671..555S,2008MNRAS.383.1241P}.  Furthermore, in S147 a strong offset between warm and cold dust can be seen and the warm dust is mainly located at the center of the S147 bubble, implying that S147 is more evolved compared to S148 and S149.

 \renewcommand{\tabcolsep}{3.0pt} 
\begin{table*}
  \caption{Physical Parameters of the \hii regions and associated massive stars.} 
\small
\centering
\label{tab:ionizing_flux} 
\begin{tabular}{cccccccccccc}
\hline \multicolumn{1}{c}{\hii} & \multicolumn{1}{c}{Star} & \multicolumn{1}{c}{$\alpha_{(2000)}$} & \multicolumn{1}{c}{$\delta_{(2000)}$} & \multicolumn{1}{c}{Flux} & \multicolumn{1}{c}{Diameter} & \multicolumn{1}{c}{$\log$ ($N_{Ly}$)} & \multicolumn{1}{c}{$T_e$} & \multicolumn{1}{c}{$n_e$} & \multicolumn{1}{c}{Pressure} & \multicolumn{1}{c}{Spectral Type}\\
\multicolumn{1}{c}{region} & \multicolumn{1}{c}{ID} & \multicolumn{1}{c}{(h:m:s)} & \multicolumn{1}{c}{(d:m:s)} & \multicolumn{1}{c}{(mJy)} & \multicolumn{1}{c}{(pc)} & \multicolumn{1}{c}{(s$^{-1}$)} & \multicolumn{1}{c}{K} & \multicolumn{1}{c}{(cm$^{-3}$)} & \multicolumn{1}{c}{($\times$10$^{-10}$ dyne cm$^{-2}$)} & \multicolumn{1}{c}{}\\
\hline

S148 &  S01  & 22:56:17.17 & +58:31:17.98 & 575.0$\pm$19.4 & 1.36  & 47.88 & 9166 & 220 & 5.60 & O9-09.5 V\\ 
S149 &  S02  & 22:56:26.52 & +58:32:12.17 & 7.1$\pm$ 0.6 & 0.50    & 45.86 & 9340 & 115 & 3.00  & B0.5-B1 V\\ 
\hline\end{tabular}
\end{table*}

\begin{figure*}
 \centering
 \includegraphics[width=14.0cm, height=9.0cm, bb=0 0 863 542]{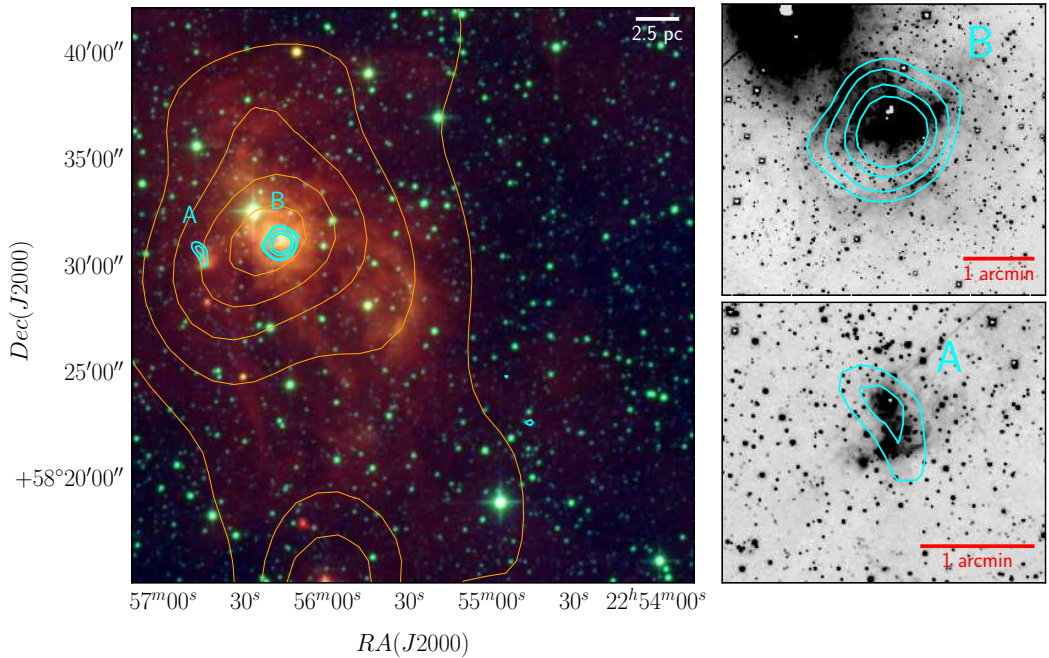}
 \caption{({\it left panel}) The surface density map of sources detected in $K$-band towards PG108.3 region overlaid on the color-composite (Red: WISE W3; Green: WISE W1; Blue: 2MASS $K$) image. Stellar surface density was calculated using the five nearest neighbors.  The {\it Planck} cold contour levels have the same meaning as in Figure \ref{fig:planck}. ({\it right panel}) The (zoomed) surface density contour of cluster `A' and `B' are overplotted on WIRCam $K$ band image.}
 \label{fig:K_density} 
\end{figure*}

\subsection{Embedded clusters and young stellar population in PG108.3 cloud complex}

\subsubsection{Stellar surface density map} \label{sec:surface_density}
Since PG108.3 cloud complex displays prominent H$\alpha$ emission at various locations (see Figure \ref{fig:halpha_spectra} left panel), a significant sub-clustering is expected  due to the fact that the massive stars are often associated to young clusters \citep[e.g.,][]{1998A&AS..133...81T,2012MNRAS.424.2486J}. We probed the sub-clusters within the molecular cloud by stellar surface density map derived using nearest neighborhood technique \citep[][]{2005ApJ...632..397G}. By using the total number of $K$ band detection within the completeness limit, we generated the stellar density profile towards PG108.3 following the method introduced by \citet[][]{1985ApJ...298...80C}. Briefly, the stellar density $\sigma(i, j)$ inside a cell of uniform grid centred at (i, j) is
$\sigma(i, j)$ = $N-1 \over \pi d^2_N(i, j)$
where $d_N$ is the distance to the N$^{th}$ nearest source from the centre of the cell. The value of N is allowed to vary depending upon our interest on the smallest scale structures of the field.  In Figure~\ref{fig:K_density} ({\it left panel}), the isodensity contours are plotted above 3$\sigma$ from  background stellar density. In Figure~\ref{fig:K_density} ({\it right panel}), the surface density contours are overlaid on the zoomed WIRCam $K$-band image. This map was obtained for nearest neighbor N = 5 in a grid size of 10$\arcsec$ $\times$ 10$\arcsec$.

The surface density in the region reveals two prominent substructures. We observed a circular structure ``B'' (peak at $\alpha_{2000}$ = $22^{\rm h}$ 56$^{\rm m}$ 16$^{\rm s}$, $\delta_{2000}$ = +58\degr 30\arcmin 59\arcsec) associated with S148 \hii region. The radius of the cluster was estimated to be $\sim$ 1.41$\arcmin$ (1.35 pc at cluster distance) from the contours above 3$\sigma$ of the background surface density. This cluster peaks nearly at the core of S148 and is extended upto \hii region boundary (see section \ref{sec:Properties_of_ionized_gas}). The second-overdensity is a cashew nut structure ``A'' (peak at $\alpha_{2000}$ = $22^{\rm h}$ 56$^{\rm m}$ 46$^{\rm s}$, $\delta_{2000}$ = +58\degr 30\arcmin 26\arcsec) associated with faint H$\alpha$ nebulosity in the eastern part of S148 \hii region.  This subcluster is extended over an effective radius 0.8$\arcmin$ (ellipticity parameters, a = 1.3$\arcmin$ and b = 0.6$\arcmin$). It is to be noted that cluster ``A'' corresponds to most massive clump ``C1'' of \citet[][]{2011AJ....141..123A}, whereas cluster ``B'' is associated with another massive clump ``C5''(see also section \ref{sec:understanding_of_star_and_cluster_formation}).

\subsubsection{Young Stellar Population}\label{young_star_search}

\begin{figure*}
 \centering
\includegraphics[width=16cm, angle=0]{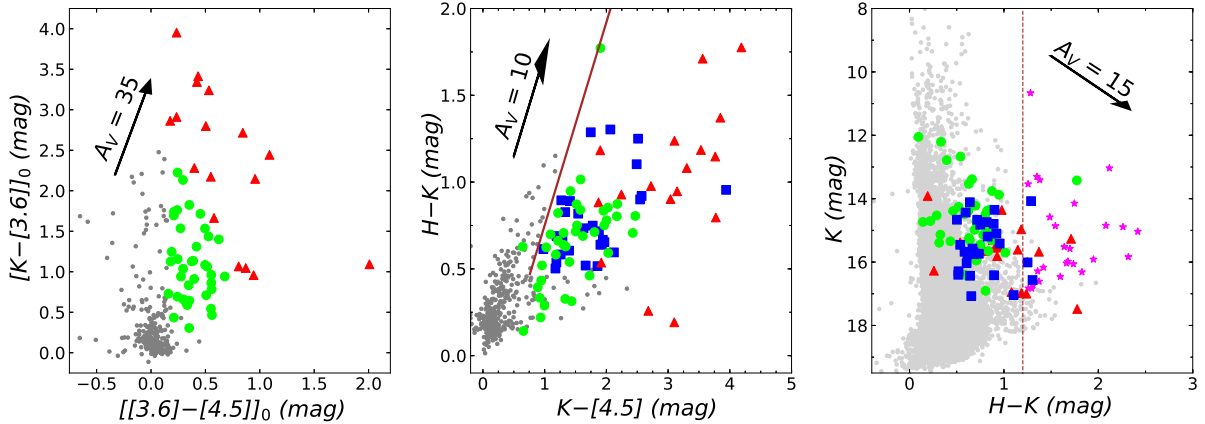}
  \caption{{\it Left}: $[3.6-4.5]_0$/$[K-3.6]_0$ CC diagram of all the uncontaminated sources identified within the {\it Spitzer} observed area. The YSOs classified as candidate class I and class II, based on \citet[][]{2009ApJS..184...18G} colour criteria, are shown in red triangles and green circles, respectively. {\it Middle}: All the detected sources are shown in $K-[4.5]$/$H-K$ diagram. The blue sources are additional YSOs detected in this method. The brown line indicates the reddening vector drawn from the tip of main-sequence. {\it Right}: $H-K/K$ CMD for all the sources. The magenta asterisk is the additional YSOs detected in this scheme. The colour cut-off at $H-K$ = 1.2 mag is shown by dashed line. The reddening vectors are  plotted for various color combinations.}  
  \label{fig:ccspitzer1}  
\end{figure*}

Crucial to our young star search is the ability to distinguish YSOs from the background-foreground stars. The IR excess emission from circumstellar material makes them distinctive from other groups of stars, although dust in the molecular cloud core averts YSOs observable at radio, submillimeter, and infrared wavelengths. We utilized $Spitzer$-IRAC, deep NIR data to distinguish IR excess sources. Note that WIRCam provides $\sim$ 3 magnitude deeper than 2MASS. Hence our analyses yield embedded sources at relatively large cluster distance of $\sim$ 3.3 kpc. The YSOs selection is described as follows:

1. Following \citet[][]{2008ApJ...674..336G,2009ApJS..184...18G}, we identified YSOs using $H$, $K$, and  3.6, 4.5 $\mu$m photometric data. First of all, we dereddened the data points using the NIR extinction map \citep[for details on the creation of the extinction map, see ][]{2016ApJ...822...49J}. The inclusion of extragalactic contamination was minimized by applying simple brightness cut in the dereddened {\it Spitzer}-IRAC photometry. All the class I and class II YSOs essentially have $[3.6]_{0}$ $\leq$ 15 mag and $[3.6]_{0}$ $\leq$ 14.5 mag, respectively. After removing the possible contaminants, we plotted all the sources in ($[[3.6]-[4.5]]_{0}$/ $[[K]-[3.6]]_{0}$) color-color (CC) diagram as shown in Figure \ref{fig:ccspitzer1} ({\it left panel}). Using colour criteria described in  \citet[][]{2009ApJS..184...18G}, we identified 55 candidate YSOs including 16 relatively high excess class I and 39 class II with intermediate IR excess emission. In Figure \ref{fig:ccspitzer1}, the class I and class II sources are displayed with the red triangles and green circles, respectively. A reddening vector $A_K$ = 2 mag is also plotted by using the reddening law from \citet[][]{2007ApJ...663.1069F}. However, we note that our present YSO selection is incomplete as sources in nebulous region might not have detected in low-sensitive {\it Spitzer} observations, particularly in the {\it Spitzer} 3.6 $\mu$m band as it is affected by poly-aromatic hydrocarbon (PAH) emissions which are often bright in the vicinity of young \hii regions \citep[][]{2010MNRAS.406..952S}.

2.  We next used $K-[4.5]$/$H-K$ CC diagram to identify additional YSOs since 4.5 $\mu$m is less affected by PAH emissions \citep[][]{2014A&A...566A.122S,2016ApJ...822...49J}. In Figure \ref{fig:ccspitzer1}({\it middle panel}), we plotted all the detected sources along with the 55 previously identified YSOs from $[[3.6]-[4.5]]_{0}$/ $[[K]-[3.6]]_{0}$. The reddening vector from the tip of the dwarf locus \citep[e.g.,][]{2006ApJ...651..502P} is also shown. Except one, all the previously identified YSOs are located at the right side of the reddening vector. A few other stars also share the same IR space with YSOs. These sources could be IR excess sources, but they were not distinguished in the above classification scheme likely due to the combination of improper dereddening and low-sensitivity at 3.6 $\mu$m. Hence, the stars above the tip of the dwarf locus ($H-K$ $>$ 0.47) and towards the right side of the reddening vectors are considered as additional excess sources in the region. This colour cut-off was determined from the comparison of the $K-[4.5]$/$H-K$ colour-colour distribution of the candidate YSOs with that of the control field region, which shows that majority of the selected YSOs are likely above $H-K$ $>$ 0.47 \citep[see][for details]{2016ApJ...822...49J}. Thus, we added 28 more stars to the YSO list.

3. Our Spitzer datasets are limited by spatial coverage (see Figure 1), while our deep NIR observations cover a larger area, therefore we used the ($H-K/K$)  colour-magnitude diagram (CMD; see the right panel of Figure \ref{fig:ccspitzer1}) to identify additional YSOs \citep[e.g.,][]{2016A&A...592A..77B}. However, trying to disentangle young stars from reddened background stars using NIR CMD  is complex, so we adopted the following approach: i) we selected brighter sources corresponding to the brightness limit of $K-[4.5]$/$H-K$ diagram, ii) we applied a $H-K$ color cut-off ($H-K$ $>$ 1.2 mag) for a source to be a YSO. The colour cut of 1.2 magnitudes is based on the maximum reddening of the field sources in the $K-[4.5]$/$H-K$ diagram. While in this approach we may miss faint YSOs and YSOs with low NIR excess, but we mainly aim to minimize the contamination to our sample. We also check a nearby reference field (centred at RA = 344.07008, Dec = 58.516937 for the same area as the cluster `B'). Using this approach, we selected another 28 YSOs in our studied region.

In total, we identified 111 YSOs in the WIRCam FOV with IR excess emission. Of these 16 have excess consistent with Class I, 39 are class II, and another 56 are termed as NIR-excess sources. The list of YSOs is presented in Table \ref{tab:yso_catalog}. We did not report Class III sources since in the present catalog they are indistinguishable from field stars based on IR excess. The YSOs selected in this method may contain unresolved galaxy contamination.  We visually inspected the high resolution $HK$ images and rejected all most-likely extended unresolved non-YSO objects in the selected YSO list. However, if we further accept that our YSOs sample is still contaminated by the non-YSO sample based on control field selected zones, then it should be less than 10\%. We suggest that spectroscopic observations are necessary for the confirmation of their membership.

\subsubsection{Mass distribution of YSOs}
\begin{figure}
 \centering
\includegraphics[width=8cm]{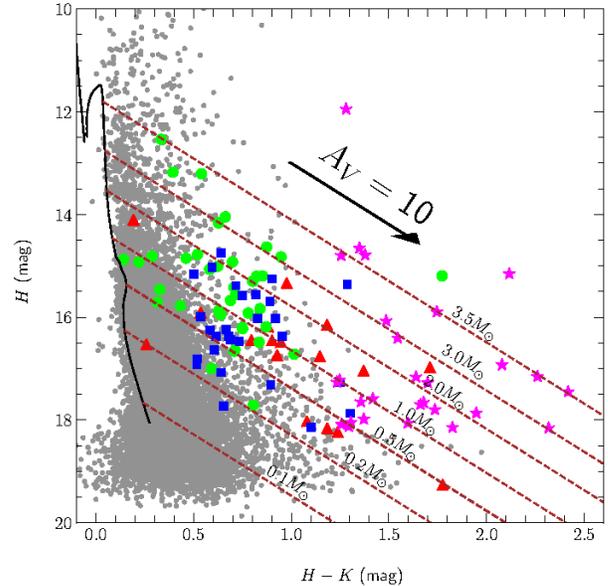}
  \caption{Mass distribution of YSOs in $H-K$/$H$ CMD. The points at various colours have the same meaning as that of Figure \ref{fig:ccspitzer1}. The curve denotes the theoretical isochrone of age 1 Myr, adopted from \citet[][]{2012MNRAS.427..127B}.}  
  \label{fig:hhk_mass}  
\end{figure}

To understand the mass sensitivity of our YSO sample, we used $H-K$/$H$ CMD as shown in Figure \ref{fig:hhk_mass}. We preferred this analysis for mass estimation since the most number of stars are detected in WIRCam $HK$ observations, on the contrary, the absence of deep MIR observations and dust obscurity of YSOs in optical wavelengths. We used the isochrone for 1 Myr as the age representative age of the region.  All the observed data points and the theoretical models were converted into CIT system using the relations given by \citet[][]{2001AJ....121.2851C}. The black curves in Figure \ref{fig:hhk_mass} is adopted from theoretical isochrone by \citet[][]{2012MNRAS.427..127B}, which is corrected for an average cluster distance 3.3 kpc. The reddening vectors for corresponding masses with slope $A_H$/$A_V$ = 0.155 and $A_K$/$A_V$ = 0.090 are taken from \citet[][]{1981ApJ...249..481C}, the typical representations of the reddening (Av = 10 mag) is also shown. Our observations allow us to prevail quite low mass end ( $\sim$ 0.1 M$\sun$). Nevertheless, our YSO detection is limited by the depth of $Spitzer$ 4.5 $\mu$m, a majority of them are $>$ 0.5 M$\sun$.  Robust investigation of the low-mass content of the clouds requires imaging with both high sensitivity and angular resolution. Nonetheless,  with a sensitivity of $\sim$ 0.2 M$\sun$, we have identified 109 YSOs, suggesting a significant number of faint YSOs might be present in the cloud assuming in young clusters the peak  of the initial mass function (IMF) lies somewhere  between 0.1-0.5 M$\sun$ \citep[e.g.,][]{2015A&A...576A.110N,2017ApJ...836...98J}.

\subsubsection{Luminosity Function and mass of the clusters associated with S148} \label{sec:K_band_luminosity} 
The $K$ band luminosity function (KLF) is a very momentous tool to diagnose the mass function and the star formation history of a young cluster \citep[][]{1993prpl.conf..429Z,2003ARA&A..41...57L}. Theoretical models on the KLF explains its dependency on several factors like the age of clusters, the sequence of star formation over the cluster, and the choice of theoretical PMS evolutionary models.  Here, we discuss the properties of the rich cluster  ``B'' of radius $\sim$ 1.41$\arcmin$ associated to S148 assuming that the maximum stars would have age $\sim$ 1 Myr.

\begin{figure}
 \centering
\includegraphics[height=6.0cm, width=8cm]{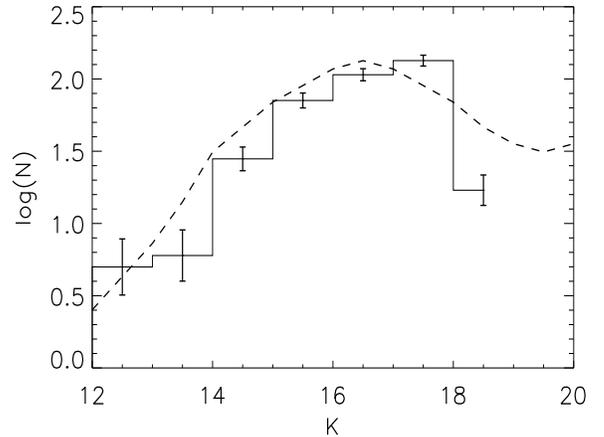}
 \caption{The $K$-band luminosity function of stars within the cluster radius after field star decontamination.  The dashed curve represents the KLF of Trapezium cluster from \citet[][]{2002ApJ...573..366M} shifted for the cluster distance of 3.3 kpc and scaled to match the peak of the KLF of S148.}
  \label{fig:K_luminosity}  
\end{figure}

We derived the KLF resulting from field star decontamination in $H-K/H$ colour-magnitude diagram (CMD) of the cluster \citep[see][for details of the field star decontamination]{2017ApJ...836...98J}. Figure \ref{fig:K_luminosity} reflects the KLF of cluster members. The KLFs of cluster rise rapidly from $K$ $\sim$ 13 to 17 mag. The KLF of S148 resembles other nearby young clusters NGC 6611 \citep[][]{2009MNRAS.392.1034O}, NGC 2024 \citep[][]{2015A&A...581A.140S}, Stock 8 \citep[e.g.,][]{2017ApJ...836...98J} etc. In Figure \ref{fig:K_luminosity}, we overplot KLF of Trapezium cluster from \citet[][]{2002ApJ...573..366M} shifted for S148 cluster distance of 3.3 kpc and also scaled to the maximum number of stars per magnitude bin between two clusters.  The photometric completeness in $K$-band is down to $K$ = 18.50 mag (see section \ref{sec:observations}) and hence the declining of the KLF below this magnitude level is not reliable.  However, the peak and flattening of the KLF is well within the completeness limit and is comparable to that of Trapezium cluster.

\begin{figure}
 \centering
\includegraphics[height=8.0cm, width=8.0cm]{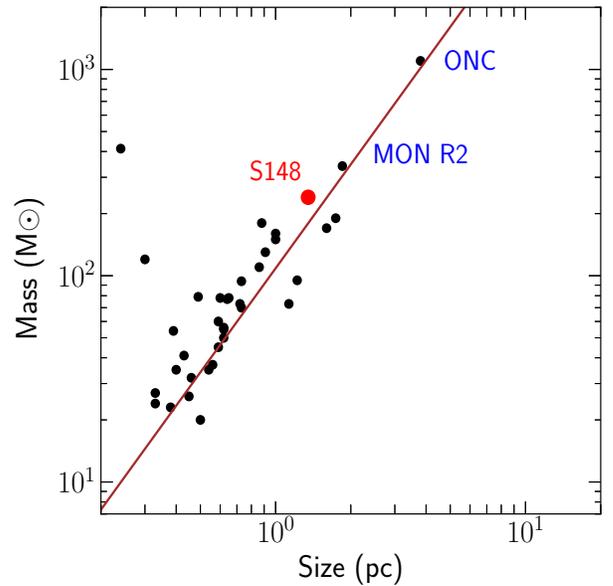}
 \caption{Comparison of S148 with the other nearby cluster sample adopted from \citet[][]{2003ARA&A..41...57L}.  The locations of S148, ONC and MON R2 are marked. The straight line is the linear fit of the sample and has a slope of $\gamma$ = 1.71 $\pm$ 0.2 obtained by \citet[][]{2016A&A...586A..68P}.}  
  \label{fig:size_mass}  
\end{figure}

Estimation of the cluster mass has to rely on the statistics of high- as well as low-mass population. It is difficult to acquire accurate information of the stars down to brown dwarf (BD) limits from present data set. However, their impact on cluster mass estimation could be neglected assuming a small contribution due to the declining slope of IMF in the BD regime. The masses of stars within the cluster area is computed from the $H-K/H$ CMD. We obtained the stellar count in different mass bins from theoretical isochrone of \citet[][]{2012MNRAS.427..127B} within the completeness limit. Using the mass-magnitude relation, we obtained the masses of the field-decontaminated stars. We integrated the mass function within the mass range 8 $-$ 0.08 M$\sun$, and the total mass of the cluster is calculated to be $\sim$ 240 M$\sun$ within the 1.41$\arcmin$ radius of the cluster. It is worthwhile to compare the cluster mass/size with the known survey of clusters  \citep[e.g., Table 1 of ][]{2003ARA&A..41...57L}. \citet[][]{2016A&A...586A..68P} obtained  a linear relation of mass and size of the clusters (Mc = CR$^\gamma$, where $\gamma$ = 1.71 $\pm$ 0.2), which provides a constraint to the theory of cluster formation. In Figure \ref{fig:size_mass}, the solid line depicts the linear relation of mass and size of the known clusters taken from \citet[][]{2003ARA&A..41...57L}.  The S148 cluster appears to be massive than several nearby young clusters and follows well the linear relation like other well known clusters MON R2 ($\sim$ 340 M$\sun$) and ONC ($\sim$ 1100 M$\sun$).

\begin{figure}
 \centering
 \includegraphics[height=7cm, width=7cm]{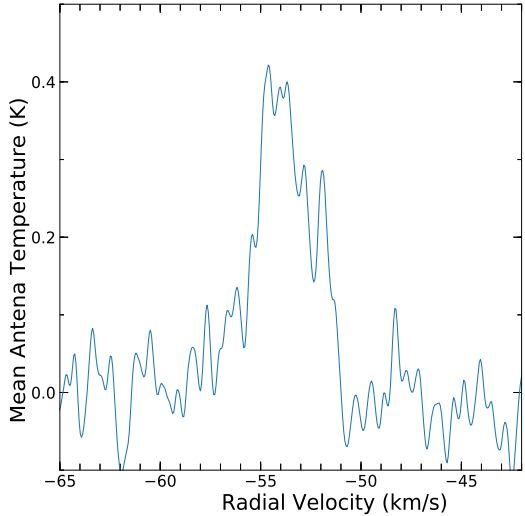}
   \caption{An integrated spectra averaged over whole studied region is displayed.}  
  \label{fig:velocity_total} 
\end{figure}

\begin{figure*}
 \centering
\includegraphics[height=16cm, width=16cm]{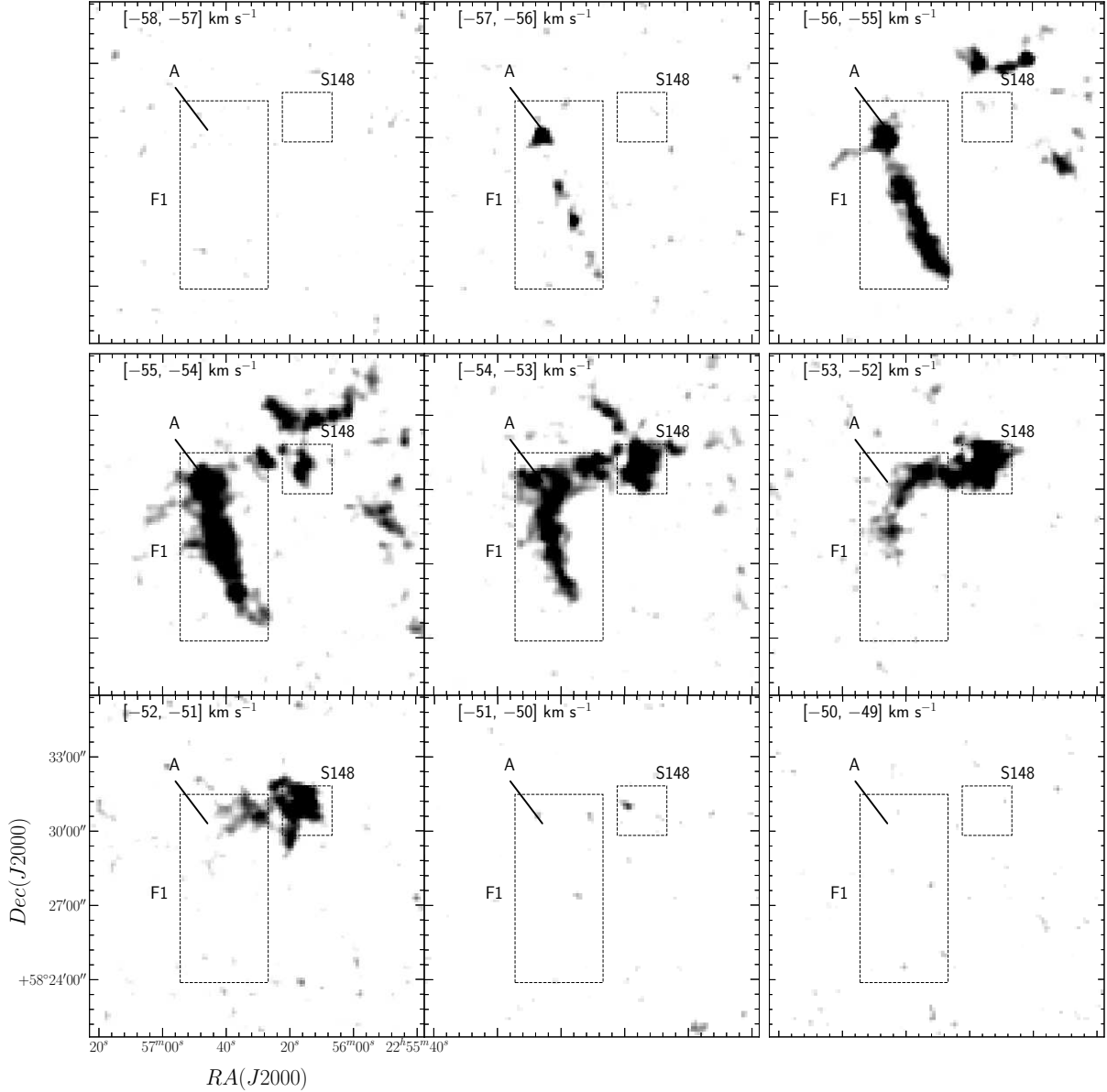}
    \caption{ The $^{13}$CO(3$-$2) velocity channel maps in the direction of the cloud. The molecular emission is integrated within the velocity range given in each  panel. The locations of filament F1 and S148 are marked. The filamentary head `A' is also mentioned.}  
  \label{fig:velocity_chennel_map} 
\end{figure*}

\begin{figure*}
\centering{
 \includegraphics[height=6.0cm, width=6.5cm]{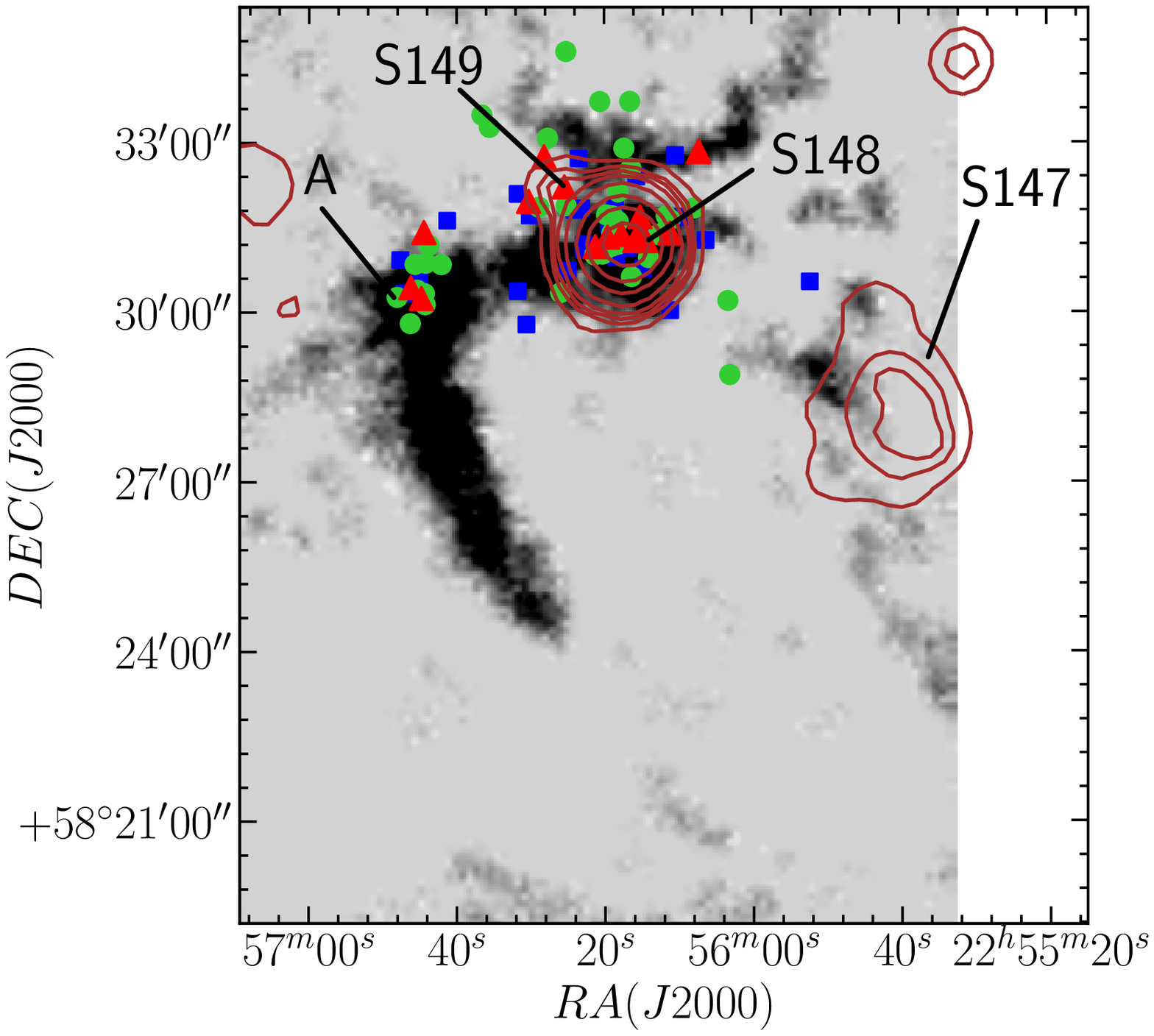}
 \includegraphics[height=6cm, width=10cm]{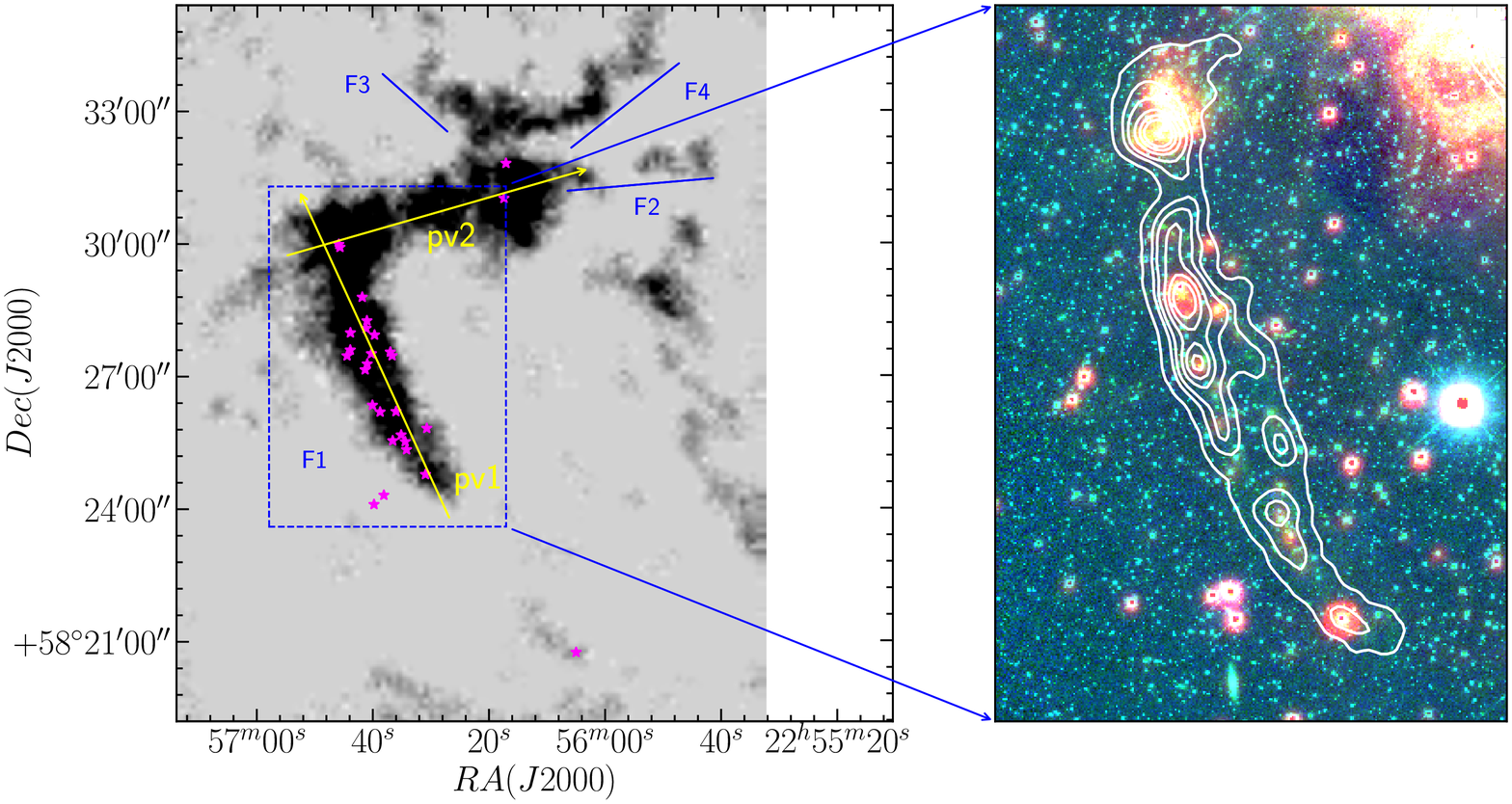}}
 \caption{({\it Left}) Spatial distribution of YSOs in PG108.3 cloud complex overplotted on the $^{13}$CO(3$-$2) image, integrated over $-$56 to $-$52 km s$^{-1}$. The points with various colours have the same meaning as that of Figure \ref{fig:ccspitzer1}. The main sub-regions are marked. The 1.4 GHz contour (brown) have the same meaning of Figure \ref{fig:iras}. ({\it Middle}) Spatial distribution of additional YSOs identified from $H-K$/$K$ CMD are shown in magenta `asterisk'. Four filaments (F1, F2, F3 \& F4) are marked in blue. Two yellow arrows (pv1 \& pv2) indicate the axes of position-velocity diagram (see Figure \ref{fig:position-velocity_diagram}). ({\it Right}) Colour-composite image of the zoomed in section of the vertical part of inverse L-shaped filament F1 (R: WISE 12 $\mu$m; G: WIRCam $K$; B: WIRCam $H$) is shown. The white contours are SCUBA 850 $\mu$m emission at 15, 50, 100, 150, 200, 250 mJy beam$^{-1}$.} 
\label{fig:spatial:ysos:integrated} 
 \end{figure*}

\subsection{Distribution of molecular gas around G108}
In the following, we study the large-scale cloud around the PG108.3 region and its correlation with the identified YSOs to understand the role of the  surrounding environment in the star formation process.

\subsubsection{Kinematics of molecular gas: evidence for the longitudinal collapse of filaments}
We used $^{13}$CO(3$-$2) observations from the JCMT to assess its large-scale kinematics and morphology. In Figure \ref{fig:velocity_total}, an average spectrum is displayed, which was obtained by averaging our studied area covering spatially the filament, cluster `A', cluster `B' (S148) and S149. While Figure \ref{fig:velocity_chennel_map} shows the channel maps in the velocity range $-$58 to $-$49 km s$^{-1}$ at intervals of 1 km s$^{-1}$. In the channel maps, on a large-scale, at least one prominent filamentary structure and several small-scale structures are visible. In particular, one can find all the structures in the channel map corresponding to the velocity  $\sim$ $-$54 km s$^{-1}$. 
Figure \ref{fig:spatial:ysos:integrated} (left \& middle panel) shows the integrated intensity map of the region for the velocity range $-$56 to $-$52 km s$^{-1}$. In the map, a potential `hub' is located in the cluster S148; where at least four filamentary structures are attached (marked in the middle panel of Figure \ref{fig:spatial:ysos:integrated}). The main cloud is clearly visible as an inverted L-shaped filamentary structure F1. Other small-scale filamentary structures are also seen near the location of S148-S149 and are marked as F2, F3, F4 in the middle panel of Figure \ref{fig:spatial:ysos:integrated}. 

\begin{figure*}
 \centering
 \includegraphics[height=6cm, width=14cm]{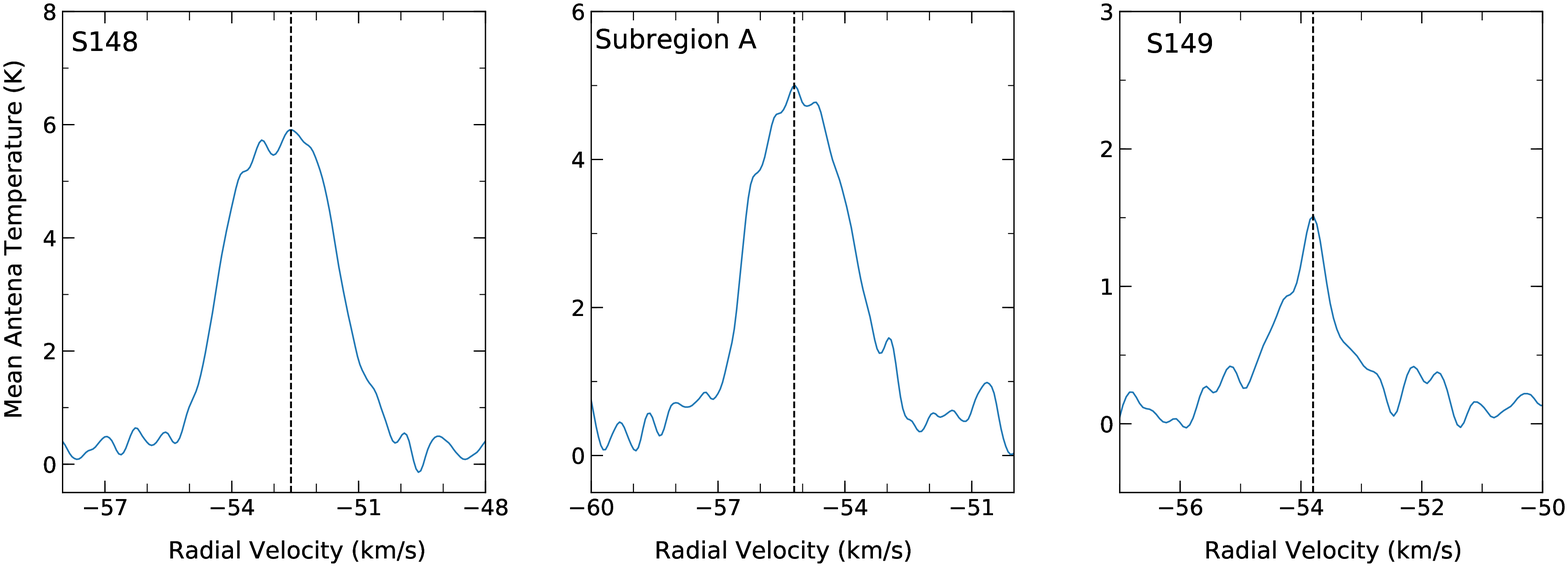}
  \caption{ The $^{13}$CO(3$-$2) spectra towards the each subregion are shown. The velocity peaks are highlighted by dashed lines.}  
  \label{fig:velocity_individual}  
\end{figure*}

In the channel map, we find that there is an overall velocity gradient in the inverted L shaped filament ranging from $-$57  km s$^{-1}$ to -51 km s$^{-1}$, from the southern end to the location of S148-S149. We also find that the gas at the southern end  of the inverted L-shaped filament is more diffuse and of lower intensity but overall the intensity distributions appear brighter as we progress closer to the location of cluster A. The peak velocity of molecular gas around cluster A and cluster B (around S148) are at $\sim$ $-$55.2 km s$^{-1}$ and $-$52.7  km s$^{-1}$, respectively (see Figure \ref{fig:velocity_individual}).

\begin{figure*}
 \centering
\includegraphics[height=7cm, width=7cm]{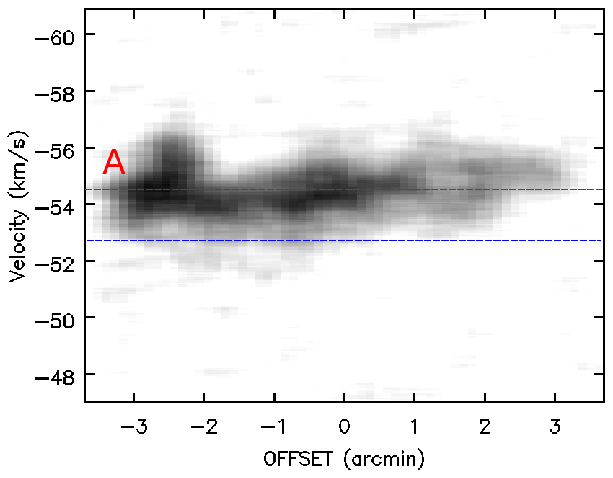}
\includegraphics[height=7cm, width=7cm]{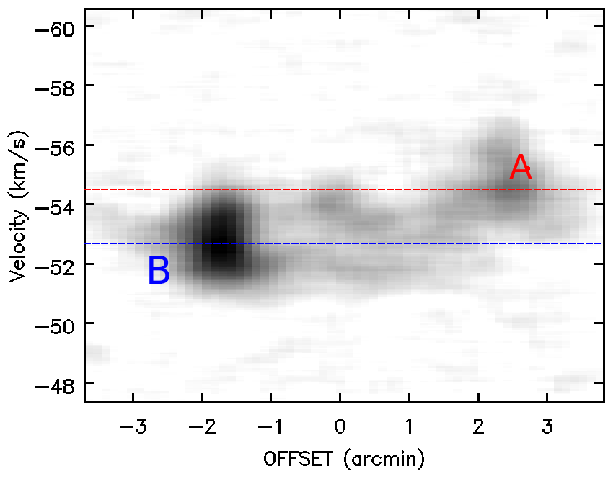}
    \caption{({\it left panel}) A position-velocity diagram along the axis ``pv1'' (along the vertical part of inverse L-shaped filament F1) as shown in the middle panel of Figure \ref{fig:spatial:ysos:integrated}. ({\it right panel}) A position-velocity diagram along the axis ``pv2'' (along the horizontal part of  inverse L-shaped filament F1) as shown in the middle panel of Figure \ref{fig:spatial:ysos:integrated}. The peak velocities of molecular gas corresponding to cluster A and cluster B (around S148) are marked by red and blue lines, respectively.  The clusters (A \& B) are marked in their corresponding positions and velocities.}  
  \label{fig:position-velocity_diagram} 
\end{figure*}

Figure \ref{fig:position-velocity_diagram} shows the position-velocity (PV) diagrams of the CO emission of the filament, where the left panel and right panel represents the PV map from the southern end of the filament (F1) to cluster A (along pv1 axes in the middle panel of  Figure \ref{fig:spatial:ysos:integrated}) and from the cluster A to S148 (along pv2 axes in the middle panel of  Figure \ref{fig:spatial:ysos:integrated}), respectively. As can be seen, the molecular material along the filaments display complex structure, however, the signature of the overall velocity gradient along the filament can be seen. The complex velocity distribution could be due to small-scale star formation within the filament since the local star formation within a filament can affect the velocity structure at local-scale due to accretion and/or feedback effects such as from outflows \citep[e.g.,][]{2015ApJ...804..141Z}. We find the velocity gradient from the southern end to cluster A and from cluster A to S148 as 0.4 and 0.6 km s$^{-1}$ pc$^{-1}$, respectively, indicating an acceleration of gas motion towards the massive clump and cluster located at the location of S148. We note other explanations such as cloud-cloud collision, feedback effects of expanding \hii regions \citep[e.g.,][]{2018ApJ...859..151L} can also generate such velocity pattern. However, we discard such hypotheses since, to create a filament of size $\sim$ 6.7  pc (length of the vertical part of the inverse L-shaped filament F1), one would expect bubbles or \hii regions of size $>$ 6.7 pc with its ionizing front parallel to the long axis of the filament. No such large bubbles or \hii regions have been observed in our studied area or in its vicinity. It is very unlikely that the compact \hii region S148 is responsible for the velocity pattern of the whole filament. On the other hand, the cloud-cloud collision enhances the density in the interface, where the massive filament can form. To form a filament of size 6.7 pc, we expect two diffuse clouds of size at least greater than 6.7 should undergo collision process. 
As shown by \citet[][]{2013MNRAS.431.3470H}, broad bridge feature connecting two intensity peaks, spatially correlated but separated in velocity, is a signature of the cloud-cloud collision. From the low-resolution Five College Radio Astronomy Observatory (FCRAO) CO Survey data of the Outer Galaxy at $^{12}$CO(1$-$0) \citep[beam $\sim$ 45$\arcsec$, velocity resolution $\sim$ 0.98 km/s;][]{1998ApJS..115..241H}, we do not observe any such two different clouds that would enhance the density at their intersection point. Velocity structure  similar to PG108.3 has been observed in other SFRs as well \citep[e.g.,][]{2014A&A...561A..83P,2017A&A...606A.123H,2017ARep...61..760K} and was interpreted as an evidence for the longitudinal collapse of filaments.
For example, \citet[][]{2014A&A...561A..83P}, suggested a global dynamical evolution of a supercritical filament initially at rest, with uniform density can be described in two stages. In the first stage, the gas at the filament-end is accelerated more efficiently, and the filament develops a linear velocity gradient increasing from centre to end. In the second stage, as the matter is accumulated at the centre, the velocity gradient starts to reverse owing to the acceleration close to the central mass becoming larger than at the filament end. Given the fact that we are witnessing a near-infrared/optical cluster at around end of the L-shaped filament and filament shows relatively high-velocity gradient near the cluster, the filament probably representing the second stage of filamentary evolution in the scheme of \citet[][]{2014A&A...561A..83P}.

\subsubsection{Correlation of molecular gas with the YSOs} 
The spatial distribution of YSOs in a star-forming complex provides an excellent tracer of recent star formation \citep[e.g.,][]{2012A&A...546A..74D,2013ApJ...764..172P,2013MNRAS.432.3445J}. Figure \ref{fig:spatial:ysos:integrated}  shows the distribution of YSOs on the integrated intensity map of CO gas, where the YSOs identified using {\it Spitzer}-IRAC and $HK$-bands are shown in the left panel, while the YSOs identified based on only $HK$ bands with $H-K$ color $>$ 1.2 mag are shown in the middle panel. As can be seen,  most of the YSOs are distributed in the close proximity of the CO emission along the filamentary cloud with enhanced concentration towards the locations of clusters A and B\footnote{We note that the close proximity of the {\it Spitzer} based YSOs with the CO emission could be due to small spatial coverage of {\it Spitzer} observations. However, we emphasize that with $K$-band surface density map in our studied area, we observed two clustering of point sources that are mimicking the location of enhanced concentrations found, implying perhaps no further young clusters/groups are present in the cloud.}. One can also see the YSOs distributed in the vertical part of F1 filament do not show any preferential clustering, rather distributed more or less linearly along the long axis. Since the YSOs identified using $HK$ bands can be biased due to high extinction often observed in filamentary environments \citep[e.g.,][]{2013ApJ...764L..26B,2015A&A...582A...1D}, in the right panel of Figure \ref{fig:spatial:ysos:integrated}, we present a zoomed in view of the vertical part of the filament F1 in the NIR and WISE 12 $\mu$m band. As can be seen, the filament contains several 12 $\mu$m sources (red sources) along its length, indicating the YSOs in the filament based on $HK$ are most likely genuine YSOs. To further confirm the nature of the red sources, we did YSOs analysis using WISE observation following the prescription given by \citet[][]{2017ApJ...845...21K}. However, WISE observations are of low-sensitivity compared to the {\it Spitzer} observations, we identified only six YSOs in our studied area, three of which lie in the vertical part of the F1 filament. We find these three sources, are already been classified as YSOs based on  $H-K$/$K$  CMD. This affirms that the most, if not all, the YSOs identified in the filament should be genuine YSOs. Such a non clustered distribution of YSOs or cores along the long axis of filaments have been observed in many filamentary systems \citep[e.g.,][]{2002ApJ...578..914H,2011A&A...533A..34H,2015A&A...581A...5S,2016A&A...588A.104G}. We note that the non-detection of the $HK$ YSOs in the other parts of the cloud could be a selection effect as we selected sources with high $H-K$ colour to minimize the effect of cloud extinction in the YSO selection. As 0.1 mag is  the mean intrinsic $H-K$ color of PMS stars of age around 1 Myr, thus a source with the $H-K$ color 1.2 mag would require a foreground cloud of visual extinction (A$_V$)  17.5 mag \citep[A$_V$ = 15.9 $\times$ E(H$-$K),][]{1985ApJ...288..618R} to be a reddened field star, implying any NIR YSO with a small excess embedded in a cloud of foreground A$_V$ less than 17.5 mag would not have been identified in our scheme.

\section{Star formation processes in pG108.3} \label{sec:understanding_of_star_and_cluster_formation}
Feedback from massive stars plays a critical role in the star formation process and evolution of molecular clouds. In particular, expanding \hii
regions may have a positive effect on star formation, i.e. they can trigger a new generation of star formation in molecular clouds either
by sweeping ambient clouds into dense shells or by compressing nearby dense clouds into bound clumps/cores \citep[for details, see][]{2010A&A...523A...6D}. In both cases, dense material eventually fragments to form new stars.  Since the PG108.3 cloud consists of three evolved \hii regions, thus one can speculate that the formation of young sources in the region might have been triggered by the expanding bubble. However, our observations largely do not favor such a process because in such a scenario one would expect the distribution of young YSOs or star-forming cores at the outskirts of the \hii regions \citep[for example, see][]{ 2006A&A...446..171Z,2013MNRAS.432.3445J,2014MNRAS.443.1614P,2016ApJ...818...95L,samal2018}. In contrast, we find no young Class I sources around S147 and most of Class I sources around S148 are co-spatial with the Class II sources located near the centre of the ionized gas. Since young clusters often harbor Class III to Class I sources as a part of cluster formation process \citep[e.g.,][]{2017ApJ...836...98J,2017MNRAS.468.2684P,2018AJ....155...44P},  thus the Class I YSOs of the S148 are most likely part of the central cluster responsible for the ionization of S148. Given the fact that the \hii region S149 located near the boundary of S148 and smaller in size, one can argue that S149 might have been influenced by S148. Unfortunately, from the current data, we have no way to estimate precisely either the age of the \hii regions or the embedded YSOs in S148. Nonetheless, we searched for signatures of the early phases of high-mass star formation such as masers and outflows in the vicinity of S148 and our search resulted in no such observations. The small size of the S149 \hii regions could be due to the fact it is ionized by less massive star compared to S148, as a result, S149 is expanding at a slow rate compared to S148 assuming both of them formed in a similar environment.  Considering the fact that S149 is optically visible, its dynamical age is close to the age of S148 and the ionized gas pressure of both regions are within a factor of two, we regard the probability that the formation of S149 is solely due to expansion of S148 rather low, although we can not ignore the possibility that the expanding ionization front of S148 might have accelerated star formation in S149 after the initial clump formation and fragmentation.  A way to prove such a hypothesis is to compare the molecular gas pressure of S149 with external gas pressure generated by the S148 \citep[e.g.,][]{2004A&A...414.1017T,2017ApJS..231....9K,2017ApJ...849...25L} in the early stage of star formation, however, in S149, neither we find significant cold gas nor \citet[][]{2011AJ....141..123A} any molecular clumps to obtain some clue in this direction.

In the PG108.3 cloud complex, we have identified two sites of cluster formation using deep $K$-band surface density map. One of them is associated with the \hii region S148 (i.e. cluster B) and is rich in infrared point sources with a total stellar mass around 240 M$_\odot$ (see \ref{sec:K_band_luminosity} for details), while the other cluster (i.e. cluster A) appears to be a loose stellar aggregate. We note, among the 13 dense compact clumps identified by \citet[][]{2011AJ....141..123A} in the vicinity of PG108.3, the most massive ($\sim$ 486 M$_\odot$)\footnote{ \citet[][]{2011AJ....141..123A} estimated the  masses of the clumps C1 and C5 as 1400 M$_\odot$ and 960 M$_\odot$, respectively, at a distance 5.6 kpc, since mass is proportional to the square of distance; we have scaled the mass at modified distance of 3.3 kpc. \label{note:mass_azimlu_distance}} clump C1 identified by them corresponds to cluster A, while the cluster B is associated to a less massive ($\sim$ 333 M$_\odot$)\footref{note:mass_azimlu_distance} clump C5. Given the fact that extinction around cluster B as estimated from the spectroscopy of the ionizing sources is $\sim$ 4 mag, implying the cluster is in the advanced stage of evolution, therefore some of the gaseous mass has already been converted into stars and some of it has been ionized by the \hii region. Moreover, the excitation temperature of $^{12}$CO(2$-$1) for C1 ($\sim$ 32 K) is colder compared to C5 ($\sim$ 40 K) clump \citep[][]{2011AJ....141..123A}, implying the star formation in clump C1 is in earlier phase compared to C5. Although, we searched for massive sources around cluster B using optical observations, but failed to identify them, likely due to high extinction. Nonetheless, in the  WISE 12 $\mu$m band, we find that the cluster shows diffuse emission at 12 $\mu$m. The 12 $\mu$m WISE band contains 11.7  $\mu$m emission commonly attributed to PAH molecule, therefore, a good tracer of newly formed, embedded B-type star formation \citep[e.g.,][]{2004ApJ...613..986P}  as these stars have the ability to heat the surrounding dust to enough temperature that can excite PAH molecules. We suggest the clump C1 is also a site of massive star formation like C5 albeit a less massive one.

Recent {\it Herschel} observations have shown that filaments and filamentary structures play a vital role in the star formation process \citep[][]{2010A&A...518L.102A}, including the formation of massive stars \citep[][]{2012A&A...540L..11S,2017arXiv170600118M}. In particular, {\it Herschel} results have emphasized that filaments are the main evolutionary stage that sets the initial condition for the formation of dense cores and stars, where the gas can be funneled along the filaments and feed the star-forming regions located at the bottom of their potential well and can facilitate the formation of star clusters. For example, in the DR21 region, several low-density striations or sub-filaments were observed perpendicular to the main filament of the complex, and they were apparently feeding matter to the main filament resulting cluster formation at the junction \citep[e.g.,][]{2010A&A...520A..49S,2012A&A...543L...3H}. In the PG108.3 complex, as can be seen in the integrated intensity $^{13}$CO(3$-$2) map shown in the left panel of Figure \ref{fig:spatial:ysos:integrated}, at least four filamentary (F1, F2, F3 \& F4) structures are attached to the location of cluster B. This is the location, where we observed a relatively rich cluster around an O9 V massive star, consistent with the picture of hub filamentary system \citep[][]{2009ApJ...700.1609M}, where several fan-like filaments can intersect, merge and fuel the clump/core located at their geometric centre to form high-mass stars and star clusters. Such hub-potential systems where the massive star and cluster formation is occurring have also been noticed in a few Galactic clouds \citep[e.g.,][]{2011A&A...527A.135C,2012ApJ...751...68L,2013ApJ...779..121G,2014A&A...561A..83P}. Although, the spatial and velocity resolution of our data is not enough to investigate the kinematics of  all the sub-filaments around the cluster A, nonetheless, from the channel maps discussed in Figure \ref{fig:velocity_chennel_map}, we observed gas motion along the F1 filament axis running all the way from its southern end to the location of clump C5. We note, while cluster formation can occur at the junction of filamentary flows, sub-filaments themselves can fragment to form stars along their length due to some kind of filament fragmentation process \citep[e.g.,][]{1997ApJ...480..681I,2018ApJ...855...24P}. These dense cores often aligned with large-scale filaments like pearls in a string and can pull matter form the global accretion flows of the system to their local potential well \citep[e.g.,][]{2015ApJ...804..141Z}. The key to understanding the fragmentation process and to distinguish between different models, such as a pure thermal fragmentation or turbulent fragmentation, require high-resolution dust as well as molecular observations \citep[e.g. see ][and references therein]{2018ApJ...855...24P}. Nonetheless, the distribution of YSOs along the F1 filament as shown in the middle panel of Figure \ref{fig:spatial:ysos:integrated}, indicates that the filament has undergone fragmentation. The most massive clump is located at one end of the L-shaped filament. We find velocity gradient along the filament, likely result from an overall inflow of mass to the center of the potential, which in this case is the location of the cluster and clump associated to S148. However as discussed earlier, PG108.3 as a whole has a more complex geometry and density structure than the ones of a single filament. The cluster is attached to a few small-filaments along with the inverted L-shaped filament more like a hub-filament system.  In such a configuration, the potential well is dominated by the central mass of the hub. However, it is difficult to conclude,  whether the small-scale filaments (F2, F3, F4) were present  before the formation of the cluster  or they have formed after the formation of the massive clump by the focusing effect of gravity of the clump that might have pulled and channeled gaseous matter around its vicinity to its center. High-resolution 
observations would shed more light on this issue. Nonetheless, it appears that the large-scale longitudinal collapse of the L-shaped filament is most likely the dominant contributor to the cluster formation.

 \renewcommand{\tabcolsep}{3.0pt} 
\begin{table*}
\caption{Catalog of YSOs towards PG108.3 cloud complex. The complete table is available in the electronic version.}
\small
\centering
\label{tab:yso_catalog}
\begin{tabular}{cccccccccccc}
\hline \multicolumn{1}{c}{RA (J2000)} & \multicolumn{1}{c}{Dec (J2000)} & \multicolumn{1}{c}{$J$} & \multicolumn{1}{c}{$eJ$} & \multicolumn{1}{c}{$H$} & \multicolumn{1}{c}{$eH$} & \multicolumn{1}{c}{$K$} & \multicolumn{1}{c}{$eK$} & \multicolumn{1}{c}{[3.6]} &\multicolumn{1}{c}{e[3.6]} & \multicolumn{1}{c}{[4.5]} & \multicolumn{1}{c}{e[4.5]}\\ 
\multicolumn{1}{c}{deg} & \multicolumn{1}{c}{(deg)} & \multicolumn{1}{c}{(mag)} & \multicolumn{1}{c}{(mag)} & \multicolumn{1}{c}{(mag)} & \multicolumn{1}{c}{(mag)} & \multicolumn{1}{c}{(mag)} & \multicolumn{1}{c}{(mag)} & \multicolumn{1}{c}{(mag)} & \multicolumn{1}{c}{(mag)} & \multicolumn{1}{c}{(mag)}& \multicolumn{1}{c}{(mag)} \\ \hline
\multicolumn{12}{c}{Class I sources}\\
\hline
344.02996 & 58.54746 & $...$ & $...$ & 16.978 &  0.017 & 15.267 &  0.020 & 12.551 &  0.010 & 11.707 &  0.008\\
344.11737 & 58.54593 & 14.279 &  0.175 & 14.108 &  0.028 & 13.916 &  0.016 & 12.824 &  0.015 & 10.816 &  0.006\\
344.05988 & 58.52124 & 16.244 &  0.190 & 15.334 &  0.025 & 14.357 &  0.022 & 12.184 &  0.007 & 11.634 &  0.008\\
344.10590 & 58.53710 & $...$ & $...$ & 16.740 &  0.039 & 15.812 &  0.026 & 14.148 &  0.035 & 13.569 &  0.035\\
344.18661 & 58.50423 & $...$ & $...$ & 16.451 &  0.024 & 15.549 &  0.015 & 12.686 &  0.022 & 12.510 &  0.030\\
344.08853 & 58.51944 & $...$ & $...$ & 16.192 &  0.023 & 15.309 &  0.021 & 14.243 &  0.033 & 13.437 &  0.030\\
344.08157 & 58.52227 & $...$ & $...$ & 15.917 &  0.015 & 15.381 &  0.019 & 14.335 &  0.039 & 13.464 &  0.030\\
344.18518 & 58.52365 & $...$ & $...$ & 18.030 &  0.041 & 16.949 &  0.018 & 14.150 &  0.061 & 13.646 &  0.070\\
344.04636 & 58.52343 & $...$ & $...$ & 19.256 &  0.098 & 17.480 &  0.027 & 13.528 &  0.018 & 13.293 &  0.024\\
344.06329 & 58.52825 & $...$ & $...$ & 16.150 &  0.022 & 14.967 &  0.023 & 14.009 &  0.029 & 13.066 &  0.022\\
\hline
\end{tabular}
\end{table*}

\section{Summary and Conclusions} \label{sec:conclusions}
In this paper, we carried out an extensive study using multi-wavelength datasets to explore the physical conditions of the molecular cloud and star formation activity around  PG108.3 cloud complex. Our analyses focused on the massive ionizing sources and ionized gas, associated young clusters, molecular gas, filamentary structure, and embedded young stellar population. We summarize the important findings of the present analyses as follows:

\begin{itemize}
\item From our optical spectroscopic analysis, we identified that \hii regions associated with the PG108.3 cloud harbor at least two massive ionizing sources. The S148 \hii region is powered by the star S01 (O9 V), whereas S149 is from the star S02 (B1 V). 

\item Two compact radio continuum sources (i.e., S01 and S02) are traced with the NVSS 1.4 GHz data, and thus our analyses reveal the \hii regions are excited by corresponding massive sources. The dynamical age of S148 was estimated to be 0.5$-$0.75 Myr.

\item The stellar surface density map reveals that at least two prominent subclusters (`A' and `B') in the vicinity of the   
PG108.3 cloud complex. The $K$-band luminosity function of S148 resembles that of Trapezium cluster. The total mass of S148 cluster is estimated to be $\sim$ 240 M$\sun$. The cluster follows the size-mass linear relation like other nearby clusters given in \citet[][]{2003ARA&A..41...57L}.

\item Using the {\it Spitzer}-IRAC and deep NIR observations, we identified 111 candidate YSOs, which include 16 Class I, 39 Class II. Our YSO estimation is sensitive to $\sim$ 0.2 M$\sun$.

\item The molecular cloud is depicted using JCMT $^{13}$CO(3$-$2) observations in the velocity range $-$57 to $-$51 km s$^{-1}$. A careful inspection of the molecular line data shows that the peak velocity of molecular gas around cluster A and cluster B (around S148) are at $-$55.2 km s$^{-1}$ and -52.7  km s$^{-1}$. The molecular data reveals the filament F1, where we found a gas motion from its southern end to S148 \hii regions via junction head at subcluster `A'. By comparing the distribution of ionized, molecular, and young stellar content, our findings do not point any triggering star formation at the \hii region boundaries, we rather suggest that the cluster formation is most likely due to the longitudinal collapse of the filament to the center of the potential well.
\end{itemize}

\acknowledgments
We thank the anonymous referee for the constructive comments which have helped to improve the scientific content and the presentation of the paper.
This work is supported by Satyendra Nath Bose National Centre for Basic Sciences, Kolkata, India under Department of Science and Technology (DST), Govt. of India. The authors are thankful to Dr Wei-Hao Wang for valuable suggestions on SIMPLE Imaging and Mosaicking PipeLinE for WIRCAM. I would like to thank Dr Ramkrishna Das for valuable suggestions. This work has made use of data from the European Space Agency (ESA) mission {\it Gaia} (\url{https://www.cosmos.esa.int/gaia}), processed by the {\it Gaia} Data Processing and Analysis Consortium (DPAC, \url{https://www.cosmos.esa.int/web/gaia/dpac/consortium}). Funding for the DPAC has been provided by national institutions, in particular the institutions
participating in the {\it Gaia} Multilateral Agreement. This research used the facilities of the Canadian Astronomy Data  Centre operated by the National Research Council of Canada with the support of the Canadian Space Agency. This publication also used the data products from the SIMBAD database (operated at CDS, Strasbourg, France), the Two Micron All Sky Survey, which is a joint project of the University of Massachusetts and the Infrared Processing and Analysis Center/California Institute of Technology, funded by NASA and NSF, archival data obtained with the Wide-Field Infrared Survey Explorer (a joint project of the University of California, Los Angeles, and the Jet Propulsion Laboratory [JPL], California Institute of Technology
[Caltech], funded by the National Aeronautics and Space Administration [NASA]), and the NOAO Science archive, which is operated by the Association of Universities for  Research in Astronomy (AURA), Inc., under a cooperative agreement with the National Science Foundation.

\software{Python, IDL, Astropy \citep[][]{2013A&A...558A..33A}, CASA \citep[][]{2007ASPC..376..127M},  DAOFIND \citep[][]{1987PASP...99..191S}, IRAF \citep[][]{1986SPIE..627..733T,1993ASPC...52..173T},  DAOPHOT  \citep[][]{1992ASPC...25..297S}, SIMPLE \citep[][]{2010ApJS..187..251W}, MOPEX (v18.5.0)}

\bibliography{s149}

\end{document}